\documentclass[a4paper,12pt]{article}

\usepackage[hmargin=.7in,vmargin=1.1in]{geometry}
\usepackage{indentfirst}
\usepackage{amsfonts}
\usepackage{mathrsfs}
\usepackage{amsmath}
\usepackage{amssymb}
\usepackage{authblk}
\usepackage{cite}
\usepackage{xcolor}
\usepackage{mathtools}
\usepackage{tensor}
\usepackage{physics}
\usepackage{graphicx}
\usepackage{bm}
\usepackage{tikz}
\usepackage{upgreek}
\usepackage{braket}
\usepackage{color,soul}
\usepackage{csquotes}
\usepackage{caption}
\usepackage{subcaption}
\usepackage{tabularx}
\usepackage{simplewick}
\usepackage[section]{placeins}
\usepackage{enumerate}
\usepackage{comment}
\newcommand{\pp}{\uppi}

\newcommand\mi{\mathrm{i}}
\newcommand\me{\mathrm{e}}

\newcommand{\dif}{\mathrm{d}}

\DeclareMathOperator{\diag}{diag}

\usepackage[bookmarksnumbered=true,bookmarksopen=true]{hyperref}
\hypersetup{colorlinks,
            linkcolor=[rgb]{0,0.3,0.6},
            citecolor=[rgb]{0,0.3,0.6},
            urlcolor=[rgb]{0,0.3,0.6}}

\title{\Large Dynamical thin shells in black hole ringdown}

\author[1,2]{Han-Wen Hu\thanks{huhanwen@itp.ac.cn}}
\author[3]{Chen Lan\thanks{stlanchen@126.com}}
\author[1,2,4]{Zong-Kuan Guo\thanks{guozk@itp.ac.cn}}

\affil[1]{\normalsize{\em Institute of Theoretical Physics, Chinese Academy of Sciences, P.O. Box 2735, Beijing 100190, China}}
\affil[2]{\normalsize{\em School of Physical Sciences, University of Chinese Academy of Sciences, No.19A Yuquan Road, Beijing 100049, China}}
\affil[3]{\normalsize{\em Department of Physics, Yantai University, 30 Qingquan Road, Yantai 264005, China}}
\affil[4]{\normalsize{\em School of Fundamental Physics and Mathematical Sciences, Hangzhou Institute for Advanced Study, University of Chinese Academy of Sciences, Hangzhou 310024, China}}

\numberwithin{equation}{section}
\date{}

\begin{document}

\maketitle

\begin{abstract}
We study how a radially oscillating thin shell outside a black hole (BH) affects gravitational wave ringdown and derive matching conditions for polar and axial perturbations from the linearized Israel junction conditions on the moving worldtube.
Periodic shell motion produces Floquet sidebands in the scattered wave.
For a distant shell, the motion has little effect on the prompt ringdown and only weakly modifies the late time echoes. When the shell is closer to the BH, a replica of a long lived shell mode enters the incident frequency band, increasing the late time echo amplitude by roughly two orders of magnitude.
We also identify an exceptional point where two polar modes coalesce and find the contribution of its second order pole in the late time waveform.
\end{abstract}

\section{Introduction}\label{sec:I}

Black hole (BH) ringdown following a merger probes the region of strong gravity.
In classical general relativity, a perturbed isolated BH settles toward a stationary Kerr BH, with its ringdown described by a set of quasinormal modes (QNMs) whose complex frequencies depend only on the BH's mass and spin.
This property underlies BH spectroscopy and tests of the no hair theorem \cite{Vishveshwara:1970cc,Teukolsky:1973ha,Leaver:1985ax,Kokkotas:1999bd,Berti:2009kk,Dreyer:2003bv,Berti:2005ys,Xia:2025hwt,Quan:2026awx}.
As gravitational wave (GW) observations with high signal to noise ratios accumulate, ringdown provides increasingly useful tests of general relativity and searches for new physics \cite{LIGOScientific:2016aoc,LIGOScientific:2016lio,LIGOScientific:2018mvr,LIGOScientific:2019fpa,LIGOScientific:2020ibl,LIGOScientific:2020tif,KAGRA:2021vkt,LIGOScientific:2021sio,Isi:2019aib,Giesler:2019uxc,Brito:2018rfr,Xia:2025hwt,Quan:2026awx}.
Astrophysical BHs, however, do not reside in ideal vacuum.
They may be surrounded by accretion disks or dark matter halos, and may also be affected by a cosmological background \cite{Shakura:1972te,Barausse:2014tra,Gondolo:1999ef,Eda:2013gg,Kavanagh:2020cfn,Konoplya:2019sns,Kiselev:2002dx,Tian:2025uvk}.
Structures outside the photon sphere can alter the effective potential and boundary conditions for BH perturbations, thereby affecting the late time waveform \cite{Leung:1997was,Konoplya:2018yrp,Jaramillo:2020tuu,Cheung:2021bol,Jaramillo:2021tmt,Berti:2022xfj,Ding:2026smj}.

Recent studies show that the BH QNM spectrum is sensitive to small perturbations in the far zone.
Localized changes to the potential or surrounding matter can shift BH QNMs substantially and introduce modes absent in vacuum. Pseudospectral analyses provide another measure of this sensitivity \cite{Nollert:1996rf,Daghigh:2020jyk,Jaramillo:2020tuu,Destounis:2021lum,Cheung:2021bol,Hu:2025efp}.
Spectral sensitivity does not necessarily produce a comparably large change in the waveform \cite{Nollert:1998ys,Jaramillo:2022kuv,Berti:2022xfj,Hu:2026sxj}.
The prompt ringdown, governed mainly by the potential barrier near the photon sphere, is usually stable. Spectral shifts, echoes, and long lived modes associated with the distant environment tend to appear in the late time response \cite{Cardoso:2016rao,Mark:2017dnq,Cardoso:2019rvt,Konoplya:2018yrp,Berti:2022xfj,Hu:2025beh,Tian:2025uvk,Hu:2026kzk}.
Two QNMs can also coalesce in both complex frequency and eigenfunction at an exceptional point (EP).
Branch migration, resonant excitation, and the corresponding time domain waveforms have been studied for Kerr BHs and perturbed effective potential models \cite{Cavalcante:2024swt,Motohashi:2024fwt,Oshita:2025ibu,Yang:2025dbn,PanossoMacedo:2025xnf,Cao:2025afs,Hu:2025efp}.
Whether these spectral changes can be distinguished in GWs depends on their effect on the waveform \cite{Vallisneri:2007ev,Lindblom:2008cm,Baibhav:2017jhs,Baibhav:2023clw,Brito:2018rfr,Barausse:2014tra}.

Thin shell models describe both reflection at an interface and the response of the matter on it. The Israel junction conditions join two vacuum spacetime regions, with surface energy density, pressure, and an equation of state assigned to the shell \cite{Israel:1966rt,Poisson:1995sv,Mazur:2001fv,Visser:2003ge}.
Studies of gravastars and other proposed alternatives to BHs have shown that such an interface changes the boundary conditions for gravitational perturbations, producing distinct QNM spectra and echo signals \cite{Chirenti:2007mk,Pani:2009ss,Pani:2010em,Chirenti:2016hzd}.
Recent analyses of a static thin shell outside a Schwarzschild BH further found little change in the main ringdown, while polar perturbations couple to the shell matter and can excite weakly damped shell modes.
On the real frequency axis, they give narrow structures associated with poles and zeros, as well as phase jumps. In the time domain, they produce weak late time echoes \cite{Laeuger:2025zgb,Torres:2022fyf,Naidoo:2021rzw,Bishop:2019ckc,Acuna-Cardenas:2024idy}.

A shell near stable equilibrium may undergo small radial oscillations, making the matching surface a moving timelike worldtube.
Linear perturbations across matching hypersurfaces, their junction conditions, and the associated gauge issues have been studied systematically \cite{Mukohyama:2000ga,Mars:2007th,Martin-Garcia:2000cgm}.
The junction conditions for the moving shell cannot be obtained by replacing the radius in the static conditions with a time dependent radius.
The shell four velocity, normal, and extrinsic curvature vary with time, while derivatives along the shell and its normal mix in the background coordinates.
Both geometrical and matter variables on the shell consequently acquire explicit time dependence.
For periodic motion, monochromatic scattering becomes a multichannel Floquet problem in which an incident wave is scattered into a discrete set of sidebands \cite{Fosco:2017jjf,Dodonov:2020eto,Zhang:2026rip}.

Here we examine the effect of a dynamical thin shell on BH perturbations.
Starting from the linearized Israel conditions on the moving worldtube, we derive the matching equations for polar and axial gravitational perturbations and use a static thin shell in a Schwarzschild background as the reference for periodic motion.
For a distant shell, motion only weakly modifies the late time echoes. The signal grows substantially, however, when the incident packet overlaps a Floquet replica of a long lived shell mode.
We also find an EP in the polar spectrum and identify its resonant contribution to the waveform.

The paper is organized as follows.
Sec.~\ref{sec:background} describes the background dynamics of the moving spherical shell, including small oscillations about equilibrium and the conservation law for shell matter.
Sec.~\ref{sec:moving_israel} derives the linearized Israel junction conditions on the moving worldtube for polar and axial perturbations.
Sec.~\ref{sec:schwarzschild} specializes these results to a Schwarzschild BH and gives the bulk master equations and the reconstruction of perturbation variables needed at the shell.
Sec.~\ref{sec:V} examines Floquet scattering, the sideband spectrum, the excitation of long lived shell modes, and the waveform signatures of EPs.
We conclude in Sec.~\ref{sec:conclusion}.

\section{Background dynamics of a moving spherical shell}\label{sec:background}

We begin with the geometry and background dynamics of an oscillating thin shell.

In units with $G=c=1$, consider a timelike thin shell hypersurface $\Sigma$ separating an exterior region $\mathcal{M}_+$ from an interior region $\mathcal{M}_-$.
The background on either side is static and spherically symmetric, with a metric function $f_{\pm}(r)$ and line element
\begin{equation}\label{eq:metric_general_f}
    \dif s_{\pm}^{2}=-f_{\pm}(r)\dif t_{\pm}^{2}+\frac{\dif r^{2}}{f_{\pm}(r)}+r^{2}\dif \Omega^{2},
\end{equation}
where $\dif\Omega^2 = \dif\theta^2 + \sin^2\theta\dif\phi^2$ is the line element of the unit sphere.

The timelike worldtube traced by the shell is parametrized by intrinsic coordinates $\xi^a = (\tau, \theta, \phi)$, with $\tau$ the proper time of an observer comoving with the shell.
Its embedding in the coordinates on either side is
\begin{equation}\label{eq:shell_embedding}
    x_{\pm}^{\mu}(\tau,\theta,\phi)=\bigl(t_{\pm}(\tau),R(\tau),\theta,\phi\bigr),
\end{equation}
where $R(\tau)$ is the shell's areal radius.
The tangent vectors to $\Sigma$ are $e^{\mu}{}_{a}=\partial x^{\mu}/\partial \xi^a$.
The induced metric $\gamma_{ab}=g_{\mu\nu}e^{\mu}{}_{a}e^{\nu}{}_{b}$ then gives $\dif s_{\Sigma}^{2}=-\dif \tau^{2}+R(\tau)^{2}\dif \Omega^{2}$.

The shell four velocity is $u_{\pm}^{\mu}=\dif x_{\pm}^{\mu}/\dif\tau$, and its unit normal is $n_{\pm}^{\mu}$.
Normalizing the four velocity, $u_{\pm}^{\mu}u_{\mu,\pm}=-1$, yields
\begin{equation}\label{eq:beta_def}
    u_{\pm}^{\mu}=\left(\frac{\beta_{\pm}}{f_{\pm}},\dot{R},0,0\right),
    \quad
    \beta_{\pm}\equiv\sqrt{\dot{R}^{2}+f_{\pm}(R)}.
\end{equation}
A dot denotes differentiation with respect to the shell proper time $\tau$.
The conditions $n_{\pm}^{\mu}n_{\mu,\pm}=1$ and $u_{\pm}^{\mu}n_{\mu,\pm}=0$ give
\begin{equation}\label{eq:normal_def}
    n_{\pm}^{\mu}=\left(\frac{\dot{R}}{f_{\pm}},\beta_{\pm},0,0\right),
    \quad
    n_{\mu,\pm}=\left(-\dot{R},\frac{\beta_{\pm}}{f_{\pm}},0,0\right).
\end{equation}
The factors $\beta_\pm$ contain both the background geometry and the shell velocity.
In the static limit $\dot R \to 0$, they reduce to the usual gravitational redshift factors $\sqrt{f_\pm}$.

The extrinsic curvature is defined by $K_{ab}^{(\pm)}=e^{\mu}{}_{a}e^{\nu}{}_{b}\nabla_{\mu}n_{\nu,\pm}$.
Spherical symmetry leaves two independent components, the temporal component $a_\pm \equiv K^\tau{}_{\tau,\pm}$ and the angular component $k_\pm \equiv K^\theta{}_{\theta,\pm}$.
Direct calculation gives
\begin{equation}\label{eq:ak_def}
    k_{\pm}\equiv K^{\theta}{}_{\theta,\pm}=\frac{\beta_{\pm}}{R},
    \quad
    a_{\pm}\equiv K^{\tau}{}_{\tau,\pm}=\frac{\ddot{R}+\frac{1}{2}f_{\pm}'(R)}{\beta_{\pm}}.
\end{equation}
A prime denotes differentiation with respect to the coordinate radius $r$.
For a shell at rest, $a_{\pm}=f_{\pm}^{\prime}/(2\sqrt{f_{\pm}})$ and $k_{\pm}=\sqrt{f_{\pm}}/R$.

To determine the shell motion, we relate these geometrical quantities to the matter on the interface.
We model the shell as an isotropic surface fluid, whose stress tensor in intrinsic coordinates is
\begin{equation}\label{eq:surface_stress}
    S^{a}{}_{b}=\diag(-\Sigma,p,p),
\end{equation}
where $\Sigma$ is the surface energy density and $p$ the tangential surface pressure.
The Israel junction conditions read
\begin{equation}\label{eq:israel_background}
    [K^{a}{}_{b}]-\delta^{a}{}_{b}[K]=-8\pp S^{a}{}_{b},
\end{equation}
where $[X]\equiv X_{+}-X_{-}$ is the jump of a quantity across the shell.
Substituting Eq.~\eqref{eq:ak_def} yields
\begin{equation}\label{eq:bg_israel_two}
    [k]=-4\pp\Sigma,
    \quad
    [a+k]=8\pp p.
\end{equation}
Using the definition of $\beta_\pm$, the first of these equations becomes
\begin{equation}\label{eq:beta_jump}
    \beta_{-}-\beta_{+}=4\pp R\Sigma \equiv \mu(R).
\end{equation}
In studies of static thin shell perturbations \cite{Pani:2009ss}, the background shell is usually fixed at $r=a$ and its nonspherical deformation treated at first order. In the present problem, the background Israel conditions first determine the trajectory $R(\tau)$, about which we introduce nonspherical perturbations.

Squaring $\beta_{-}=\beta_{+}+\mu$ and using $\beta_{\pm}^{2}=\dot{R}^{2}+f_{\pm}(R)$ gives
\begin{equation}\label{eq:veff_eq}
    \dot{R}^{2}+V_{\rm eff}(R)=0,
\end{equation}
with effective potential
\begin{equation}\label{eq:veff_def}
    V_{\rm eff}(R)=\frac{f_{+}(R)+f_{-}(R)}{2}
    -\frac{\mu(R)^{2}}{4}
    -\frac{\bigl[f_{-}(R)-f_{+}(R)\bigr]^{2}}{4\mu(R)^{2}}.
\end{equation}
The radial dynamics depends on both metric functions $f_\pm$ and the shell mass function $\mu(R)$.
Similar worldlines of oscillating thin shells can arise in gravitational collapse models \cite{Gao:2008jy,Dokuchaev:2010fv,Huang:2021bdm}.
An equilibrium radius $R_0$ satisfies
\begin{equation}\label{eq:equilibrium_conditions}
    V_{\rm eff}(R_0)=0, \quad V_{\rm eff}^{\prime}(R_0)=0.
\end{equation}
Stable oscillations require a local minimum of the effective potential, $V_{\rm eff}^{\prime\prime}(R_0)>0$.
These conditions describe a stationary shell with zero oscillation amplitude.
Small accretion or an external gravitational disturbance can displace a physical shell from its static equilibrium radius.
In the first integral description, such a displacement corresponds to a nearby radial trajectory with a small integration constant, or equivalently to a small shift of the effective potential. The curvature of the potential well still sets the leading restoring force.
Near the stable point, write $R(\tau)=R_{0}+\xi(\tau)$ and retain terms linear in $\xi\sim\mathcal{O}(\epsilon)$.
Differentiating the first integral with respect to $\tau$ and linearizing around $R_0$ gives $\ddot{\xi}+\Omega_{\tau}^{2}\xi=0$, where $\Omega_{\tau}^{2}=V_{\rm eff}^{\prime\prime}(R_0)/2$.
To leading order in the radial excitation, the shell trajectory in its proper frame is therefore
\begin{equation}\label{eq:R_oscillation_tau}
    R(\tau)=R_{0}+\epsilon\cos(\Omega_{\tau}\tau+\phi_{0}),
    \quad
    \epsilon\ll R_{0}.
\end{equation}
In the nonspherical perturbation analysis below, we take Eq.~\eqref{eq:R_oscillation_tau} as a prescribed periodic background worldtube and linearize the gravitational perturbations and nonspherical shell displacement about it.
For an observer at infinity, we use the exterior static time $t_+$. The shell frequency must therefore be converted from proper time to this coordinate time.
To leading order in the oscillation amplitude, $\dif t_{+}/\dif\tau=\beta_{+}/f_{+}$ gives
\begin{equation}\label{eq:Omega_schw}
    \Omega_+=\Omega_\tau\sqrt{f_+(R_0)}.
\end{equation}

Determining the shell motion also requires $\mu(R)$. Surface energy conservation, $D_{a}S^{a}{}_{b}=0$, gives
\begin{equation}\label{eq:surface_conservation}
    \dot{\Sigma}+2\frac{\dot{R}}{R}(\Sigma+p)=0.
\end{equation}
For $\dot{R} \neq 0$, this is equivalent to $\dif\Sigma/\dif R=-2(\Sigma+p)/R$.
Near the shell equilibrium radius, we treat the pressure as a function of the density, $p(\Sigma(R))$, and expand about $\Sigma_0$ to obtain
\begin{equation}\label{eq:eos_background}
    p(R)=p_{0}+v_{s}^{2}\bigl[\Sigma(R)-\Sigma_{0}\bigr],
\end{equation}
where $v_{s}^{2}=\dif p/\dif\Sigma \big|_{R_{0}}$ is the squared sound speed of the surface fluid.

\section{Linearized Israel conditions on a moving worldtube}\label{sec:moving_israel}

The background Israel conditions yield the shell equation of motion $\dot{R}^2 + V_{\rm eff}(R) = 0$.
We now derive the matching conditions for gravitational perturbations on this moving boundary.
Because the shell oscillates about equilibrium, its perturbed surface generally departs from the background trajectory $r=R(\tau)$ by a physical normal displacement $\delta x^\mu = \Phi(\tau) Y_{\ell m} n^\mu$.
An axial gravitational perturbation does not change the shell's areal radius, so $\Phi=0$ in that sector.
Directly perturbing the moving boundary in static bulk coordinates would mix metric perturbations with terms caused by shifting the evaluation point.
We use a comoving fixed worldtube gauge and a Gaussian normal decomposition adapted to the shell. This pulls the perturbed surface back to the background worldtube and incorporates its physical displacement into an effective metric perturbation.

\subsection{Polar sector}

For polar perturbations, an infinitesimal displacement vector $\Phi n^\mu$ along the background normal pulls the deformed shell back to the trajectory $R(\tau)$ in the fixed worldtube gauge.
The total effective metric perturbation seen by the shell is
\begin{equation}\label{eq:Delta_g_def}
    \Delta g_{\mu\nu} = h_{\mu\nu} + \mathcal{L}_{\Phi n} g^{(0)}_{\mu\nu},
\end{equation}
where $h_{\mu\nu}$ is the linear gravitational perturbation in the bulk and $\mathcal{L}_{\Phi n}$ is the Lie derivative along $\Phi n^{\mu}$.
The interface conditions can then be evaluated on the unperturbed moving shell.
Because the shell tangent and normal are orthogonal, $e^\mu_a n_{\mu,\pm} = 0$, the tangential projection of the Lie derivative term in Eq.~\eqref{eq:Delta_g_def} reduces to the extrinsic curvature contribution $e^\mu_a e^\nu_b \mathcal{L}_{\Phi n} g^{(0)}_{\mu\nu} = 2\Phi Y_{\ell m}K_{ab}^{(\pm)}$.
The first order perturbation of the induced metric is therefore
\begin{equation}\label{eq:Delta_gamma_ab}
    \Delta \gamma_{ab} = h_{\mu\nu} e^\mu_a e^\nu_b + 2\Phi Y_{\ell m} K_{ab}^{(\pm)}.
\end{equation}
The second term is entirely due to the geometrical pullback.
To project the four dimensional polar perturbation $h_{\mu\nu}$ onto the shell, define
\begin{equation}\label{eq:polar_projections}
    \mathcal{U}_\pm \equiv h_{\mu\nu}^{(\pm)} u^\mu_\pm u^\nu_\pm, \quad
    \mathcal{J}_\pm \equiv h_{\mu\nu}^{(\pm)} u^\mu_\pm n^\nu_\pm, \quad
    \mathcal{N}_\pm \equiv h_{\mu\nu}^{(\pm)} n^\mu_\pm n^\nu_\pm,
\end{equation}
and the angular trace perturbation $\mathcal{K}_\pm$, defined by $h_{AB}^{(\pm)} = R^2 \mathcal{K}_\pm Y_{\ell m} \Omega_{AB}$ with $\Omega_{AB}$ the metric on the unit sphere.
Projecting Eq.~\eqref{eq:Delta_gamma_ab} onto the $(\tau,\tau)$ and $(A,B)$ components gives
\begin{subequations}
\begin{align}
    \Delta \gamma_{\tau\tau} &= (\mathcal{U}_\pm - 2a_\pm \Phi) Y_{\ell m}, \label{eq:Delta_gamma_tau_tau}\\
    \Delta \gamma_{AB} &= R^2 (\mathcal{K}_\pm + 2k_\pm \Phi) Y_{\ell m} \Omega_{AB}. \label{eq:Delta_gamma_AB}
\end{align}
\end{subequations}

The induced metric must be continuous across the shell, $[\Delta \gamma_{ab}] = 0$.
Its angular component gives $[\mathcal{K}] + 2[k]\Phi = 0$.
Using the background Israel equation $[k] = -4\pp\Sigma(R)$, we obtain the first polar matching condition,
\begin{equation}\label{eq:E1}
    [\mathcal{K}] = 8\pp\Sigma\Phi.
\end{equation}
The temporal component gives $[\mathcal{U}] - 2[a]\Phi = 0$.
Since $[a+k] = 8\pp p$ implies $[a] = 8\pp p + 4\pp\Sigma$, the second polar condition is
\begin{equation}\label{eq:E2}
    [\mathcal{U}] = 8\pp(\Sigma + 2p) \Phi.
\end{equation}
These two conditions follow from continuity of the induced metric and the shell matter distribution without relying on a particular algebraic form of the background metric.

To calculate the perturbation of the extrinsic curvature, introduce Gaussian normal coordinates $(\eta,\xi^a) \equiv (\eta,\tau,\theta,\phi)$ near the shell.
The coordinate $\eta$ runs along background normal geodesics, with the unperturbed shell at $\eta=0$.
Locally the background metric has the form
\begin{equation}\label{eq:gn_background_metric}
    \dif s^2 = \dif \eta^2 + \gamma_{ab}^{(0)}(\eta,\xi)\dif \xi^a \dif \xi^b .
\end{equation}
For a spherically symmetric background it can also be written as
\begin{equation}\label{eq:gn_background_metric_AB}
    \dif s^2 = \dif \eta^2 - A^2(\eta,\tau)\dif \tau^2 + B^2(\eta,\tau)\dif \Omega^2 .
\end{equation}
On the shell, $A(0,\tau)=1$ and $B(0,\tau)=R(\tau)$.
The functions $A(\eta,\tau)$ and $B(\eta,\tau)$ are local functions in coordinates adapted to the shell, distinct from the bulk metric functions $f_\pm(r)$.
They separate normal expansion, evolution along the shell, and motion relative to the bulk coordinates. The final matching conditions involve only shell quantities such as $f_\pm(R)$, $\beta_\pm$, $a_\pm$, and $k_\pm$.
In Gaussian normal coordinates, $n_\mu\dif x^\mu=\dif\eta$, and we decompose the effective metric perturbation as
\begin{equation}\label{eq:gn_perturbation_decomposition}
    \Delta g_{\eta\eta}=c, \quad \Delta g_{\eta a}=p_a, \quad \Delta g_{ab}=q_{ab},
\end{equation}
where $c$ changes the normal line element, $p_a$ mixes normal and tangential directions, and $q_{ab}$ perturbs the geometry tangent to the shell.
For a shell fixed at a constant coordinate radius, the mixed component $p_a$ can often be eliminated in static shell perturbation theory.
For a moving shell, it accounts for the effect of interface motion on the coordinates and connects the physical shell displacement to the bulk metric perturbation.

In the fixed worldtube gauge, the deformed shell is pulled back along the background normal to $\eta=0$ by $\xi^\mu=\Phi(\tau)Y_{\ell m}n^\mu$.
In Gaussian normal coordinates, $\xi_\eta=\Phi Y_{\ell m}$ and $\xi_a=0$.
We treat $\Phi$ as intrinsic to the shell and extend it into a small normal neighborhood by choosing $D_n\Phi=0$.
This choice specifies how the Lie derivative is evaluated near the shell without imposing a dynamical constraint on its deformation.
The final matching conditions depend on $\Phi$ and its proper time derivatives on the shell, not on how this normal extension is continued away from it.
We project $\Delta g_{\mu\nu}$ into the Gaussian normal coordinates to find the components $c$, $p_a$, and $q_{ab}$ needed in the extrinsic curvature perturbation.
In Regge-Wheeler gauge, the bulk perturbation projects to $h_{\eta\eta}=\mathcal{N}_\pm Y_{\ell m}$, $h_{\tau\eta}=\mathcal{J}_\pm Y_{\ell m}$, $h_{\tau\tau} = A^2\mathcal{U}_\pm Y_{\ell m}$, and $h_{AB}= B^2\mathcal{K}_\pm Y_{\ell m}\Omega_{AB}$ on the shell.
The Lie derivative adds no $(\eta,\eta)$ contribution because $g_{\eta\eta}=1$. Its $(\eta,a)$ component is $\partial_a(\Phi Y_{\ell m})$, and its tangential component is $2\Phi K_{ab}^{(\pm)}Y_{\ell m}$.
Consequently,
\begin{equation}\label{eq:c_p_q_components}
    c=\mathcal{N}_\pm Y_{\ell m},\quad
    p_\tau=(\mathcal{J}_\pm+\dot{\Phi})Y_{\ell m},\quad
    p_A=\Phi D_A Y_{\ell m},
\end{equation}
where $D_A$ is the covariant derivative on the unit sphere.
The remaining components are
\begin{equation}\label{eq:q_components_gn}
    q_{\tau\tau}=(\mathcal{U}_\pm-2a_\pm\Phi)Y_{\ell m},\quad
    q_{AB}=R^2(\mathcal{K}_\pm+2k_\pm\Phi)Y_{\ell m}\Omega_{AB}.
\end{equation}
The $\dot{\Phi}$ in Eq.~\eqref{eq:c_p_q_components} comes from the proper time evolution of the shell's physical displacement and is absent for a fixed static boundary.
It enters $\Delta a_\pm$, together with tangential time derivatives of the bulk perturbation, in determining the polar response of the shell matter.

For the polar extrinsic curvature conditions, write the Gaussian normal metric in ADM form along $\eta$
\begin{equation}\label{eq:adm_eta_split}
    \dif s^2=N^2\dif\eta^2+\gamma_{ab}\left(\dif\xi^a+N^a\dif\eta\right)
    \left(\dif\xi^b+N^b\dif\eta\right).
\end{equation}
The extrinsic curvature is then
\begin{equation}\label{eq:K_adm_def}
    K_{ab}=\frac{1}{2N}\left(\partial_\eta\gamma_{ab}-D_aN_b-D_bN_a\right),
\end{equation}
where $N$ is the lapse and $N_a$ the shift.
For the background shell in Gaussian normal coordinates, $N=1$ and $N_a=0$.
At first order, $\Delta g_{\eta a}=p_a$ gives $N_a=p_a$, while $\Delta g_{\eta\eta}=c$ gives $N=1+c/2$.
Substituting these relations into Eq.~\eqref{eq:K_adm_def} and retaining linear terms yields
\begin{equation}\label{eq:Delta_K_identity}
    \Delta K_{ab}^{(\pm)}=\frac{1}{2}\left(D_n q_{ab}-D_a p_b-D_b p_a-c K_{ab}^{(\pm)}\right).
\end{equation}
Because $\eta$ runs along $n^\mu$, $\partial_\eta$ is the normal derivative $D_n = n^\mu \partial_\mu$ in these coordinates. In particular, $\partial_\eta q_{ab}=D_n q_{ab}$.
The background metric on each spherical section of the shell is $\sigma_{AB}=R^2\Omega_{AB}$, so $k_\pm=\sigma^{AB}K_{AB}^{(\pm)}/2$.
Its first order perturbation contains changes in both the inverse metric and the extrinsic curvature
\begin{equation}\label{eq:Delta_k_definition}
    \Delta k_\pm=
    \frac{1}{2}\sigma^{AB}\Delta K_{AB}^{(\pm)}
    +\frac{1}{2}\Delta\sigma^{AB}K_{AB}^{(\pm)} .
\end{equation}
Equation~\eqref{eq:q_components_gn} gives $\Delta\sigma_{AB} = q_{AB} = R^2(\mathcal{K}_\pm+2k_\pm\Phi)Y_{\ell m}\Omega_{AB}$ and hence $\Delta\sigma^{AB}=-(\mathcal{K}_\pm+2k_\pm\Phi)Y_{\ell m}\sigma^{AB}$.
Since $K_{AB}^{(\pm)}=k_\pm\sigma_{AB}$ in the background, the second term in Eq.~\eqref{eq:Delta_k_definition} is $-k_\pm(\mathcal{K}_\pm+2k_\pm\Phi)Y_{\ell m}$.
The first term follows from the $(A,B)$ component of Eq.~\eqref{eq:Delta_K_identity}.
Using $D_n(R^2)=2k_\pm R^2$ and noting that neither $Y_{\ell m}$ nor $\Omega_{AB}$ varies along the normal gives
\begin{equation}\label{eq:Dn_qAB_intermediate}
    D_n q_{AB}=R^2\left(2k_\pm\mathcal{K}_\pm+4k_\pm^2\Phi
    +D_n\mathcal{K}_\pm+2\Phi D_n k_\pm \right)Y_{\ell m}\Omega_{AB}.
\end{equation}
Meanwhile, $p_A=\Phi D_A Y_{\ell m}$ gives $\sigma^{AB}(D_A p_B+D_B p_A)=-2\ell(\ell+1)\Phi Y_{\ell m}/R^2-4\frac{\dot R}{R}(\mathcal{J}_\pm+\dot{\Phi})Y_{\ell m}$.
Substituting these expressions into Eq.~\eqref{eq:Delta_k_definition}, the terms proportional to $k_\pm\mathcal{K}_\pm$ and $k_\pm^2\Phi$ cancel, leaving
\begin{equation}\label{eq:Delta_k_result}
    \Delta k_\pm=
    \left[\frac{1}{2}\left(D_n\mathcal{K}_\pm-k_\pm\mathcal{N}_\pm\right)
    +\Phi D_nk_\pm+\frac{\ell(\ell+1)}{2R^2}\Phi
    +\frac{\dot R}{R}(\mathcal{J}_\pm+\dot{\Phi})\right]Y_{\ell m}.
\end{equation}
Perturbing the background equation $[k]=-4\pp\Sigma$ gives $[\Delta k]=-4\pp\delta\Sigma$.
Substituting $k_\pm=\beta_\pm/R$ and its normal derivative yields the third polar matching condition,
\begin{equation}\label{eq:E3}
    \left[\frac{\beta}{R}\mathcal{N}-D_n\mathcal{K}\right]
    +\left[\frac{2\beta^2}{R^2}-\frac{f^{\prime}}{R}\right]\Phi
    -2\frac{\dot R}{R}[\mathcal{J}]
    =8\pp\delta\Sigma .
\end{equation}
This equation gives the geometrical origin of the surface density perturbation.
The value of $\delta\Sigma$ follows from the jump in normal derivatives of the polar gravitational perturbation and from the change in extrinsic curvature when the background shell undergoes the normal displacement $\Phi$.

Next consider the temporal component of the perturbed extrinsic curvature.
Setting $(a,b)=(\tau,\tau)$ in Eq.~\eqref{eq:Delta_K_identity} and using $p_\tau=(\mathcal{J}_\pm+\dot{\Phi})Y_{\ell m}$ and $q_{\tau\tau}=(\mathcal{U}_\pm-2a_\pm\Phi)Y_{\ell m}$ gives
\begin{equation}\label{eq:Delta_a_result}
    \Delta a_\pm=
    \left[-\frac{1}{2}D_n\mathcal{U}_\pm+D_u\mathcal{J}_\pm
    -\frac{1}{2}a_\pm\mathcal{N}_\pm+\Phi D_na_\pm+\ddot{\Phi}\right]Y_{\ell m}.
\end{equation}
The terms $D_n\mathcal{U}_\pm$, $D_u\mathcal{J}_\pm$, and $a_\pm\mathcal{N}_\pm$ describe how the bulk polar perturbation changes the shell's temporal extrinsic curvature.
The terms $\Phi D_na_\pm$ and $\ddot{\Phi}$ arise from normal motion of the shell itself.
In particular, $\ddot{\Phi}$ is the inertial response of the shell driven by the polar perturbation.
It appears in the expression for $\Delta a_\pm$ on either side but cancels when the jump is taken, since the displacement $\Phi$ is shared by the two sides.
Adding Eqs.~\eqref{eq:Delta_a_result} and \eqref{eq:Delta_k_result} and perturbing the other background Israel equation, $[a+k]=8\pp p$, gives the fourth polar matching condition
\begin{equation}\label{eq:E4}
    \left[D_n\mathcal{K}-D_n\mathcal{U}+2D_u\mathcal{J}-(a+k)\mathcal{N}\right]
    +2\left[D_n(a+k)\right]\Phi
    +2\frac{\dot R}{R}[\mathcal{J}]
    =16\pp\delta p .
\end{equation}
This equation describes the response of the tangential shell pressure to polar gravitational perturbations.
As in the background Israel equations, the condition on $[k]$ controls the surface density response, while the one on $[a+k]$ controls the tangential stress.
For a local barotropic equation of state, the first order matter perturbations obey
\begin{equation}\label{eq:surface_eos_perturbation}
    \delta p=v_s^2\delta\Sigma .
\end{equation}
We use a local, nondissipative, inviscid barotropic surface fluid and take $v_s^2$ to be constant near equilibrium.
If the matter response depends on frequency, or if the fluid dissipates energy or has additional intrinsic degrees of freedom, Eq.~\eqref{eq:surface_eos_perturbation} must be generalized. These changes would affect the polar shell modes and Floquet sideband strengths.
Together with the bulk fields, Eqs.~\eqref{eq:E3} and \eqref{eq:E4} then determine the algebraic relations involving $\Phi$ and $\delta\Sigma$.

The dynamical polar matching problem is thus fixed by continuity of the induced metric, Eqs.~\eqref{eq:E1} and \eqref{eq:E2}, and by the extrinsic curvature jumps, Eqs.~\eqref{eq:E3} and \eqref{eq:E4}.
All four conditions hold on the moving worldtube traced by the same nonstationary shell $R(\tau)$. They cannot be obtained simply by replacing a constant radius with $R(\tau)$ in the static conditions.
Relative to a fixed static shell, the directional derivatives $D_u$ and $D_n$ explicitly contain the velocity $\dot{R}$ and mix evolution along the shell with normal gradients in the bulk coordinates.
The terms $\Phi D_n k$ and $\Phi D_n(a+k)$ describe the geometrical correction to the background extrinsic curvature at the physically displaced interface.
These effects make the matching conditions time dependent and produce Floquet sidebands in the scattered wave.

\subsection{Axial sector}

The axial matching problem has fewer shell degrees of freedom than the polar one, although axial GWs still interact with the shell.
Polar perturbations change the area element of the sphere and hence perturb the surface density and pressure.
Axial perturbations instead twist the sphere without changing its areal radius at first order, so there is no normal displacement and $\Phi=0$.
We use a minimal model for axial perturbations of the shell in which the surface fluid has no independent odd parity tangential velocity perturbation or toroidal surface flow.
Axial perturbations then enter the junction conditions through projections of the bulk geometry on the two sides, without an additional surface matter degree of freedom of the $S_{\tau A}^{\rm odd}$ type.
An independent odd parity surface flow would add a matter source to these equations. We do not include such a flow.
The geometrical pullback term $\mathcal{L}_{\Phi n}g_{\mu\nu}^{(0)}$ of the fixed worldtube gauge vanishes, and the axial conditions follow directly from tangential and normal projections of the bulk perturbation onto the moving shell.

In Regge-Wheeler gauge, the only nonzero components of the odd parity metric perturbation are the $(t,A)$ and $(r,A)$ components.
We write them as
\begin{equation}\label{eq:odd_metric_components}
    \delta g_{tA}^{\rm odd}=-h_0(t,r)X_A^{\ell m},
    \quad
    \delta g_{rA}^{\rm odd}=-h_1(t,r)X_A^{\ell m},
\end{equation}
where $X_A^{\ell m}$ denotes the axial vector spherical harmonic.
Since $\Phi=0$, the shell remains on its background worldtube $R(\tau)$, although it generally moves in the bulk coordinates.
Its projections must therefore be formed with the four velocity $u_\pm^\mu$ and normal $n_\pm^\mu$ of the moving shell even when the axial perturbation has no normal displacement.
This distinction separates the dynamical junction conditions from those of a static shell.

The odd parity perturbation contributes only to the $(\tau,A)$ component of the induced metric on the shell
\begin{equation}\label{eq:odd_induced_metric_tauA}
    \delta\gamma_{\tau A}^{\rm odd}
    =
    u^\mu_\pm e_A^\nu \delta g_{\mu\nu}^{\rm odd}
    =
    u^\mu_\pm \delta g_{\mu A}^{\rm odd}.
\end{equation}
Substituting $u_\pm^\mu=(\beta_\pm/f_\pm,\dot{R},0,0)$ yields
\begin{equation}\label{eq:odd_u_projection}
    \delta\gamma_{\tau A}^{\rm odd}
    =
    -\left(\frac{\beta_\pm}{f_\pm}h_{0,\pm}+\dot{R}h_{1,\pm}\right)X_A^{\ell m}.
\end{equation}
We define the axial projection along the shell's time direction by
\begin{equation}\label{eq:frak_u_def}
    \mathfrak{u}_\pm \equiv \frac{\beta_\pm}{f_\pm}h_{0,\pm}+\dot{R}h_{1,\pm}.
\end{equation}
Continuity, $[\delta\gamma_{\tau A}^{\rm odd}]=0$, then requires
\begin{equation}\label{eq:odd_condition_u}
    [\mathfrak{u}]=0.
\end{equation}
Thus the two sides induce the same twisting tangential geometry at a given event on the shell.

Next consider the odd parity part of the extrinsic curvature.
In the minimal model for axial perturbations specified above, the surface stress tensor has no independent odd parity tangential shear source.
The linearized Israel conditions therefore set to zero the odd parity jump in the extrinsic curvature perturbation that cannot be supported by surface matter.
In the axial sector, this condition can be expressed through continuity of the normal projection,
\begin{equation}\label{eq:odd_normal_projection_raw}
    n^\mu_\pm\delta g_{\mu A}^{\rm odd} = n^t_\pm\delta g_{tA}^{\rm odd}
    +
    n^r_\pm\delta g_{rA}^{\rm odd}.
\end{equation}
Using $n_\pm^\mu=(\dot{R}/f_\pm,\beta_\pm,0,0)$ gives
\begin{equation}\label{eq:odd_normal_projection}
    n^\mu_\pm\delta g_{\mu A}^{\rm odd}
    =
    -\left(\frac{\dot{R}}{f_\pm}h_{0,\pm}+\beta_\pm h_{1,\pm}\right)X_A^{\ell m}.
\end{equation}
Define the axial projection along the normal direction as
\begin{equation}\label{eq:frak_n_def}
    \mathfrak{n}_\pm
    \equiv
    \frac{\dot{R}}{f_\pm}h_{0,\pm}+\beta_\pm h_{1,\pm}.
\end{equation}
With no independent axial surface flow, the second junction condition becomes
\begin{equation}\label{eq:odd_condition_n}
    [\mathfrak{n}]=0.
\end{equation}
Equation~\eqref{eq:odd_condition_u} constrains the axial geometry along the shell's time direction, whereas Eq.~\eqref{eq:odd_condition_n} constrains its normal projection.
Together they ensure that an axial GW crossing the shell does not create a tangential geometrical discontinuity unsupported by surface stress.

In the static limit $\dot{R}=0$, Eqs.~\eqref{eq:odd_condition_u} and \eqref{eq:odd_condition_n} reduce to $[\beta h_0/f]=0$ and $[\beta h_1]=0$, respectively, the axial conditions on a fixed shell.
For an oscillating shell, the $\dot{R}$ terms in its four velocity and normal mix $h_0$ and $h_1$ in both conditions. They arise from the moving background, rather than from an imposed boundary model.
Thus, even without a compressional shell mode, the axial sector responds to the motion of the background shell through changes in the reflection phase and echo arrival times.

\section{Schwarzschild background}\label{sec:schwarzschild}

The preceding junction conditions apply to a general spherically symmetric background.
To obtain a solvable boundary value problem, we take both bulk regions to be Schwarzschild spacetimes,
\begin{equation}\label{eq:f_schwarzschild_pm}
    f_{\pm}(r)=1-\frac{2M_{\pm}}{r}.
\end{equation}
The exterior ADM mass generally differs from the interior BH mass because the shell itself carries surface energy, so $M_+\neq M_-$.
We denote the shell mass by $\delta M\equiv M_+-M_-$.
Consequently, the two coordinate times $t_+$ and $t_-$ do not define a common global time.
The two geometries are matched at the same value of the shell's proper time $\tau$.
For a moving shell, the directional derivatives $D_u$ and $D_n$ already contain its velocity $\dot{R}$.
Every junction condition must be evaluated at the same physical event $\left(t_\pm(\tau),R(\tau),\theta,\phi\right)$ on the shell, rather than on slices of equal coordinate time.

We fix the perturbation gauge in each Schwarzschild region.
For gravitational radiation with $\ell\geq2$, let $Y_{\ell m}$ denote the scalar spherical harmonics and $D_A$ the covariant derivative on the unit sphere.
We define the vector harmonics by $Y_A^{\ell m}=D_A Y_{\ell m}$ and $X_A^{\ell m}=-\epsilon_A{}^B D_BY_{\ell m}$.
In Regge-Wheeler gauge, the metric perturbation separates into polar and axial parts,
\begin{equation}\label{eq:RWG_metric_perturbation_matrix} h_{\mu\nu,s} =
\begin{pmatrix}
f_sH_{0,s}Y & H_{1,s}Y & -h_{0,s}X_\theta & -h_{0,s} X_\phi \\
* & f_s^{-1}H_{2,s}Y & -h_{1,s}X_\theta & -h_{1,s} X_\phi \\
* & * & r^2K_sY & 0 \\
* & * & * & r^2\sin^2\theta K_s Y
\end{pmatrix},
\end{equation}
where $H_{0,s}$, $H_{1,s}$, $H_{2,s}$, and $K_s$ are the polar variables, while $h_{0,s}$ and $h_{1,s}$ are the axial variables.
The linearized Einstein equations in each Schwarzschild region impose the algebraic constraint $H_{0,s}=H_{2,s}$.
We keep these variables separate while forming the geometrical projections and impose the constraint when reconstructing the metric perturbation from the master variables.

The thin shell leaves the vacuum propagation equations on either side unchanged.
With the tortoise coordinate defined by $\dif r_{*,s}/\dif r=1/f_s(r)$, the axial master function $\Psi_s^{\rm odd}$ satisfies the Regge-Wheeler equation
\begin{equation}\label{eq:RW_equation_schw}
    \left[-\partial_{t_s}^2+\partial_{r_{*,s}}^2-V_{{\rm RW},s}(r)\right]\Psi_s^{\rm odd}=0,
\end{equation}
where
\begin{equation}\label{eq:RW_potential_schw}
    V_{{\rm RW},s}(r)=f_s(r)\left[\frac{\ell(\ell+1)}{r^2}-\frac{6M_s}{r^3}\right].
\end{equation}
The polar master function $Z_s$ obeys the Zerilli equation
\begin{equation}\label{eq:Z_equation_schw}
    \left[-\partial_{t_s}^2+\partial_{r_{*,s}}^2-V_{{\rm Z},s}(r)\right]Z_s=0.
\end{equation}
With $\lambda=(\ell-1)(\ell+2)/2$, the Zerilli potential is
\begin{equation}\label{eq:Z_potential_schw}
    V_{{\rm Z},s}(r)=
    \frac{2f_s(r)}{r^3(\lambda r+3M_s)^2}
    \left[\lambda^2(\lambda+1)r^3+3\lambda^2M_sr^2
    +9\lambda M_s^2r+9M_s^3\right].
\end{equation}
The shell couples the solutions on its two sides through the junction conditions, without altering either bulk equation.
A moving timelike interface $R(\tau)$ thus joins two otherwise separate BH perturbation problems.
Unlike a static modification of the potential throughout the bulk, the interface retains the shell's surface tension, surface density, and equation of state, together with its response to the GW.

The geometrical projections can be expressed in terms of the Schwarzschild perturbation variables.
The shell's four velocity and normal are still denoted by $u_s^\mu$ and $n_s^\mu$.
In the axial sector, $\Phi=0$.
The standard relations $h_{0,s}=f_s\partial_r(r\Psi_s^{\rm odd})/2$ and $h_{1,s}=r\partial_{t_s}\Psi_s^{\rm odd}/2f_s$ turn $[\mathfrak{u}]=0$ and $[\mathfrak{n}]=0$ into boundary equations for $\Psi_s^{\rm odd}$ of the form
\begin{equation}\label{eq:odd_master_condition_1_section}
\left[
\frac{\beta}{2}\left(R\Psi^{{\rm odd}\,\prime}+\Psi^{\rm odd}\right)
+\frac{R\dot{R}}{2f}\partial_t\Psi^{\rm odd}
\right]=0,
\end{equation}
\begin{equation}\label{eq:odd_master_condition_2_section}
\left[
\frac{\dot{R}}{2}\left(R\Psi^{{\rm odd}\,\prime}+\Psi^{\rm odd}\right)
+\frac{R\beta}{2f}\partial_t\Psi^{\rm odd}
\right]=0.
\end{equation}
In each region, the derivatives $\partial_t$ in these expressions mean $\partial_{t_s}$.
For time dependence $\me^{-\mi\omega_s t_s}$, one replaces $\partial_{t_s}$ by $-\mi\omega_s$.
The conditions reduce to those of a static shell when $\dot{R}=0$. Terms due to motion arise from the projections onto the shell itself.

In the polar sector, we first relate $H_{0,s}$, $H_{1,s}$, $H_{2,s}$, and $K_s$ to the geometrical projections introduced above.
Direct substitution of Eq.~\eqref{eq:RWG_metric_perturbation_matrix} and the definitions of $u^\mu$ and $n^\mu$ gives
\begin{subequations}\label{eq:polar_projection_metric_variables_section}
    \begin{align}
    \mathcal{U}_s
    &=
    \frac{\beta_s^2}{f_s}H_{0,s}
    +2\frac{\beta_s\dot{R}}{f_s}H_{1,s}
    +\frac{\dot{R}^2}{f_s}H_{2,s},\\
    \mathcal{N}_s
    &=
    \frac{\dot{R}^2}{f_s}H_{0,s}
    +2\frac{\beta_s\dot{R}}{f_s}H_{1,s}
    +\frac{\beta_s^2}{f_s}H_{2,s},\\
    \mathcal{J}_s
    &=
    \frac{\beta_s\dot{R}}{f_s}H_{0,s}
    +\frac{\beta_s^2+\dot{R}^2}{f_s}H_{1,s}
    +\frac{\beta_s\dot{R}}{f_s}H_{2,s},
\end{align}
and
\begin{equation}\label{eq:K_projection_metric_variable_section}
    \mathcal{K}_s=K_s.
\end{equation}
\end{subequations}
The linearized polar Israel conditions are Eqs.~\eqref{eq:E1}, \eqref{eq:E2}, \eqref{eq:E3}, and \eqref{eq:E4}.
For the shell matter we use the local barotropic equation of state, whose first order perturbation obeys Eq.~\eqref{eq:surface_eos_perturbation}.
The shell variables $\Phi$ and $\delta\Sigma$ must therefore be solved together with the Zerilli fields on both sides.
This coupling to the shell matter distinguishes polar perturbations from axial ones.

The polar metric variables can be reconstructed from the Zerilli master function.
In vacuum Schwarzschild regions, $H_{0,s}=H_{2,s}\equiv H_s$, and Regge-Wheeler gauge gives
\begin{subequations}
\begin{equation}\label{eq:Z_reconstruction_time_domain_section}
    H_s=-\mathcal{A}_sZ_s+\mathcal{B}_sZ_s^{\prime}+rf_sZ_s^{\prime\prime},
    \quad
    K_s=\mathcal{D}_sZ_s+f_sZ_s^{\prime},
\end{equation}
and
\begin{equation}\label{eq:H1_reconstruction_time_domain_section}
    H_{1,s}=\mathcal{C}_s\partial_{t}Z_s+r\partial_{t}Z_s^{\prime}.
\end{equation}
\end{subequations}
The coefficient functions $\mathcal{A}_s(r)$, $\mathcal{B}_s(r)$, $\mathcal{C}_s(r)$, and $\mathcal{D}_s(r)$ are
\begin{subequations}\label{eq:Z_reconstruction_coefficients_section}
\begin{align}
    \mathcal{A}_s(r)
    &=
    \frac{
    \lambda^2(\lambda+1)r^3+3\lambda^2M_sr^2
    +9\lambda M_s^2r+9M_s^3
    }
    {r^2(\lambda r+3M_s)^2},\\
    \mathcal{B}_s(r)
    &=
    \frac{\lambda r^2-\lambda M_sr-3M_s^2}{r(\lambda r+3M_s)},\\
    \mathcal{C}_s(r)
    &=
    \frac{\lambda r^2-3\lambda M_sr-3M_s^2}
    {r f_s(\lambda r+3M_s)},\\
    \mathcal{D}_s(r)
    &=
    \frac{\lambda(\lambda+1)r^2+3\lambda M_sr+6M_s^2}
    {r^2(\lambda r+3M_s)}.
\end{align}
\end{subequations}
In the frequency domain, Eq.~\eqref{eq:H1_reconstruction_time_domain_section} becomes
\begin{equation}\label{eq:H1_reconstruction_frequency_domain_section}
    H_{1,s}=-\mi\omega_s\left(\mathcal{C}_sZ_s+rZ_s^{\prime}\right).
\end{equation}
Equation~\eqref{eq:Z_reconstruction_time_domain_section} contains the second radial derivative $Z_s^{\prime\prime}$.
Combining $\partial_{r_{*,s}}^2Z_s=f_s^2Z_s^{\prime\prime}+f_sf_s^{\prime}Z_s^{\prime}$ with the Zerilli equation \eqref{eq:Z_equation_schw} gives
\begin{equation}\label{eq:Z_second_derivative_elimination_section}
    Z_s^{\prime\prime} = \frac{1}{f_s^2} \left(\partial_{t}^2Z_s+V_{{\rm Z},s}Z_s-f_sf_s^{\prime}Z_s^{\prime}\right).
\end{equation}
Eliminating $Z_s^{\prime\prime}$ avoids computing second radial differences at the interface. The matching procedure needs only the master function and its first radial derivative.

The static limit has $\dot R=\ddot R=0$, with the interface fixed at $R=R_0$.
Time translation symmetry is then restored. Scattering states can be labeled by the exterior frequency $\omega$, and different frequencies decouple. The time mapping at the shell gives the interior frequency $\omega_-=\omega/\alpha_0$, where $\alpha_0=\sqrt{f_+(R_0)/f_-(R_0)}$.
We consider radiation incident on the shell from the interior. The reflected wave obeys a purely ingoing condition at the BH horizon, while the exterior field is purely outgoing at infinity.
With time factors $\me^{-\mi\omega_s t_s}$ on either side, the outgoing radial wave at infinity is proportional to $\me^{+\mi\omega_+r_{*,+}}$, and the ingoing radial wave at the horizon is proportional to $\me^{-\mi\omega_-r_{*,-}}$.
The master variables can therefore be expanded as
\begin{equation}\label{eq:static_scattering_basis_section}
    \Psi_{-,\omega}=\psi_{-,\omega}^{\rm in}+B(\omega)\psi_{-,\omega}^{\rm H}, \quad \Psi_{+,\omega} = C(\omega)\psi_{+,\omega}^{\rm out}.
\end{equation}
The subscript $\omega$ labels the exterior frequency throughout. The interior radial functions are solved at the corresponding $\omega_-$. The incident amplitude is normalized to unity, and the Israel junction conditions determine $B(\omega)$ and $C(\omega)$.
For each $\omega$, we impose asymptotic conditions near the BH horizon, $r_*\to-\infty$, and at spatial infinity, $r_*\to+\infty$, then integrate the appropriate Regge-Wheeler or Zerilli equation to $r=R_0$. This gives the field values $\psi$ and first radial derivatives $\psi^{\prime}$ on both sides of the interface. We then substitute them into the Israel junction conditions.
For axial perturbations, the static limits of Eqs.~\eqref{eq:odd_master_condition_1_section} and \eqref{eq:odd_master_condition_2_section} form a $2\times2$ algebraic system for $(B(\omega),C(\omega))$.
For polar perturbations, we solve Eqs.~\eqref{eq:E1} through \eqref{eq:E4} after reconstructing $H_s$, $K_s$, and $H_{1,s}$ from $Z_s$ and $Z_s^{\prime}$.
Equation~\eqref{eq:Z_second_derivative_elimination_section} eliminates $Z_s^{\prime\prime}$ analytically. This avoids higher order numerical differences at the interface and reduces the risk of numerical error amplification.

\section{Floquet scattering response of a periodically moving shell}\label{sec:V}

For a static shell, time translation symmetry ensures that frequency is conserved during scattering.
We now consider a shell undergoing small periodic oscillations about its equilibrium radius.
As the boundary moves, the junction conditions on its worldtube become time dependent and scattering couples different frequencies.
An incident monochromatic wave then produces a discrete set of frequency components, leading naturally to a Floquet scattering problem.
The time dependence of the Floquet matrix comes from the periodic changes in the shell position, four velocity, normal, and directional derivatives $D_u$ and $D_n$. It cannot be obtained by replacing $R_0$ with $R(t)$ in the static junction conditions.

\subsection{Floquet formulation on a moving shell}

In the calculation we use the exterior Schwarzschild time $t \equiv t_+$ as the coordinate time of an observer at infinity.
Because the masses differ across the shell, the interior and exterior coordinate times run at different rates.
Matching the perturbation fields at the same physical event requires the time map $\dif t_- / \dif t = \beta_- f_+ / (\beta_+ f_-)$.
For periodic shell motion, $\dif t_- / \dif t$ is itself periodic.
Its integral, the interior time $\Theta_-(t)$, is the sum of a term linear in $t$ and a periodic part
\begin{equation}\label{eq:time_map_decomposition}
\Theta_-(t) = \alpha_- t + \chi(t) ,
\end{equation}
where $\alpha_-$ is the mean redshift factor and $\chi(t)$ is periodic with $\chi(0)=0$. These quantities are given by
\begin{equation}\label{eq:alpha_and_chi_definition}
\alpha_- = \frac{1}{T}\int_0^T \frac{\beta_- f_+}{\beta_+ f_-} \dif t , \quad \chi(t) = \int_0^t \frac{\beta_- f_+}{\beta_+ f_-} \dif t' - \alpha_- t .
\end{equation}
The function $\chi(t)$ describes the periodic propagation delay introduced by the moving shell over a cycle.
It cannot be absorbed into a constant redshift and produces additional sideband mixing when the interior phase is projected onto Floquet harmonics.
The period in exterior coordinate time is $T=\int_0^{2\pp/\Omega_\tau}(\beta_+/f_+)\dif\tau$.
We henceforth write $\Omega=2\pp/T$, so that $\alpha_-$ relates the evolution frequencies on the two sides.

The moving boundary breaks continuous time translation symmetry but preserves a discrete symmetry of period $T$.
Floquet scattering states can therefore be labeled by a quasifrequency $\varpi$.
For a given incident quasifrequency, the exterior gravitational perturbation contains sidebands at frequencies
\begin{equation}\label{eq:floquet_sideband_frequency}
\omega_{n,+} = \varpi + n\Omega , \quad n \in \mathbb{Z}.
\end{equation}
The corresponding interior and exterior frequencies satisfy
\begin{equation}\label{eq:inner_sideband_frequency}
\omega_{n,-} = \frac{\omega_{n,+}}{\alpha_-} = \frac{\varpi + n\Omega}{\alpha_-} .
\end{equation}
Away from the shell, each sideband satisfies the standard axial or polar perturbation equation in its own bulk region.
As in the static problem, we consider radiation emitted from inside the shell. An incident wave with quasifrequency $\varpi$ occupies the central channel $n=0$ and propagates toward the shell.
The exterior field is purely outgoing at infinity,
\begin{equation}\label{eq:outer_floquet_expansion}
\Psi_+(t, r) = \sum_{n=-N}^{N} C_n(\varpi) \psi^{\rm out}_{n,+}(r) \me^{-\mi \omega_{n,+} t} ,
\end{equation}
where $C_n(\varpi)$ is the transmission amplitude of output sideband $n$.
The interior field contains the prescribed monochromatic incident component and reflected sidebands directed toward the BH horizon, and can be written as
\begin{equation}\label{eq:inner_floquet_expansion}
\Psi_-(t_-, r) = \psi^{\rm in}_{0,-}(r) \me^{-\mi \omega_{0,-} t_-} + \sum_{n=-N}^{N} B_n(\varpi) \psi^{\rm H}_{n,-}(r) \me^{-\mi \omega_{n,-} t_-},
\end{equation}
where $\psi^{\rm in}_{0,-}$ is the radial function of the interior wave incident on the shell, and $\psi^{\rm H}_{n,-}$ is the radial function of a wave falling into the horizon.
On the shell, substituting $t_-=\Theta_-(t)$ brings the periodic term in Eq.~\eqref{eq:time_map_decomposition} into each sideband phase.
Sidebands do not interact in the bulk. Frequency conversion occurs at the moving shell because the coefficients of the junction conditions vary periodically in time.

Floquet projection turns the time dependent boundary problem into an algebraic system.
Substituting the perturbation fields and their derivatives on either side of the comoving worldtube into the linearized Israel conditions gives boundary equations of the form
\begin{equation}\label{eq:residual_general_definition}
    \mathcal{F}_\alpha(t) = 0 , \quad \alpha = 1, \cdots, N_{\rm c}.
\end{equation}
There are two matching conditions in the axial sector, $N_{\rm c}=2$, and four in the polar sector, $N_{\rm c}=4$.
Each boundary function is a linear combination of bulk amplitudes and intrinsic shell variables.
At fixed incident quasifrequency $\varpi$, discrete time translation symmetry gives the Floquet expansion
\begin{equation}\label{eq:residual_floquet_expansion}
    \mathcal{F}_\alpha(t) = \me^{-\mi\varpi t} \sum_{n=-\infty}^{\infty} \mathcal{F}_{\alpha n} \me^{-\mi n\Omega t} .
\end{equation}
The junction conditions hold only if every Fourier coefficient vanishes, $\mathcal{F}_{\alpha n} = 0$.
We isolate the coefficients by projecting each boundary function over one background period to obtain
\begin{equation}\label{eq:floquet_residual_projection}
\mathcal{F}_{\alpha n} = \frac{1}{T} \int_0^T \mathcal{F}_\alpha(t) \me^{\mi(\varpi + n\Omega)t} \dif t = 0 .
\end{equation}
Because the junction conditions are linear, the Floquet expansion of $\mathcal{F}_\alpha(t)$ is linear in the unknown amplitudes.
Consider an unknown $X_m$ in frequency channel $m$, such as a reflection amplitude $B_m$, a transmission amplitude $C_m$, or a shell matter perturbation.
Its contribution to a boundary equation generally has the form
\begin{equation}\label{eq:residual_single_contribution}
    \mathcal{F}_\alpha(t) \supset \mathcal{B}_{\alpha m}(t; \varpi) X_m \me^{-\mi\omega_{m,-}\Theta_-(t)} .
\end{equation}
The factor $\mathcal{B}_{\alpha m}(t;\varpi)$ varies with time through the geometrical projections generated by the shell's periodic position, velocity, and acceleration.

Substituting Eq.~\eqref{eq:residual_single_contribution} into Eq.~\eqref{eq:floquet_residual_projection} gives the matrix element that couples the $m$th unknown to the $n$th projected boundary equation as
\begin{equation}\label{eq:general_matrix_element}
\mathbb{M}_{\alpha n, m} = \frac{1}{T} \int_0^T \mathcal{B}_{\alpha m}(t; \varpi) \me^{\mi(\varpi+n\Omega)t} \me^{-\mi\omega_{m,-}\Theta_-(t)} \dif t .
\end{equation}
The matrix structure is easiest to see if we temporarily neglect $\chi(t)$ and use the mean redshift approximation $\Theta_-(t) \simeq \alpha_- t$.
Then $\omega_{m,-}\Theta_-(t) \simeq (\varpi+m\Omega)t$, and Eq.~\eqref{eq:general_matrix_element} becomes
\begin{equation}\label{eq:toeplitz_like_approximation}
\mathbb{M}_{\alpha n, m} \simeq \mathcal{B}_\alpha^{[n-m]}(\omega_m) = \frac{1}{T} \int_0^T \mathcal{B}_{\alpha m}(t; \varpi) \me^{\mi(n-m)\Omega t} \dif t .
\end{equation}
The matrix elements are thus controlled mainly by the difference $n-m$ between row and column indices.
If $\mathcal{B}_{\alpha m}$ were independent of the incident frequency, elements with the same $n-m$ would be equal, giving a Toeplitz matrix.
In the calculation, however, the bulk radial functions within $\mathcal{B}_{\alpha m}$ also depend on the absolute frequency $\omega_m$, so elements on the same diagonal are not strictly constant.
The dependence on $n-m$ gives the Floquet scattering matrix its banded structure.

After projection, the unknown sideband amplitudes form a closed linear system.
The axial shell has no normal deformation or intrinsic fluid perturbation, leaving only the bulk reflection and transmission amplitudes $B_n$ and $C_n$ in each channel.
Collecting them into $\mathbb{X}^{\mathrm{ax}} = (B_{-N}, C_{-N}, \cdots, B_N, C_N)^{\rm T}$ and denoting the incident terms by $\mathbb{B}_{\mathrm{in}}^{\mathrm{ax}}$, we obtain
\begin{equation}\label{eq:axial_matrix_equation}
\mathbb{M}^{\mathrm{ax}}(\varpi) \mathbb{X}^{\mathrm{ax}} = \mathbb{B}_{\mathrm{in}}^{\mathrm{ax}}(\varpi) .
\end{equation}
For a sideband cutoff $N$, the coefficient matrix has dimension $2(2N+1)$.
The source vector follows by substituting the known interior incident wave $\psi^{\rm in}_{0,-}$ alone into the junction conditions and Fourier projecting the resulting boundary term over the period $T$.
The incident wave is prescribed only in the central channel $n=0$, so the noncentral output sidebands arise entirely from modulation by shell motion.
In the polar sector, the shell is a dynamical system as well as a geometrical interface.
Each channel contains four unknowns: the bulk amplitudes $B_n$ and $C_n$, the shell displacement amplitude $\Phi_n$, and the density perturbation amplitude $\delta\Sigma_n$.
We collect them in $\mathbb{X}^{\mathrm{pol}} = \left(\Phi_{-N}, \delta\Sigma_{-N}, B_{-N}, C_{-N}, \cdots, \Phi_N, \delta\Sigma_N, B_N, C_N \right)^{\rm T}$.
The polar junction conditions then give
\begin{equation}\label{eq:polar_matrix_equation}
\mathbb{M}^{\mathrm{pol}}(\varpi) \mathbb{X}^{\mathrm{pol}} = \mathbb{B}_{\mathrm{in}}^{\mathrm{pol}}(\varpi) .
\end{equation}
This is a $4(2N+1) \times 4(2N+1)$ system.
Solving it shows how shell deformation and the surface equation of state couple gravitational perturbations in different frequency channels.
Numerically, we truncate the infinite Floquet system at sideband order $N$.
The coupling to the moving boundary weakens rapidly with increasing $|n|$, and convergence tests allow us to use $N=7$ for all calculations below.

We reconstruct the waveform at infinity by inverse Fourier transformation of the frequency domain amplitudes.
The incident Gaussian packet, expressed in terms of the quasifrequency $\varpi$, is
\begin{equation}\label{eq:dynamic_gaussian_packet}
A_{\rm in}(\varpi) = \exp\left[-\frac{(\varpi-\omega_c)^2}{2\sigma_\omega^2}\right] \me^{-\mi \varpi u_0} ,
\end{equation}
where $\omega_c$ is the central frequency, $\sigma_\omega$ is the spectral width, and $u_0$ shifts the packet in time. We take $\omega_c M_+=0.37$ and $\sigma_\omega M_+=0.05$ in the numerical calculations. In the $\ell=2$ channel, direct BH ringdown is mainly controlled by the fundamental mode, whose oscillation frequency satisfies $M_+\Re\omega_0\simeq0.37$. The Gaussian describes radiation of finite bandwidth propagating outward from the region near the BH to the shell.
We use $u=t_+-r_{*,+}$ for retarded time at infinity.
Summing the output channels gives the signal at infinity as
\begin{equation}\label{eq:dynamic_floquet_waveform}
h_{\rm out}(u) = \Re \int_{-\infty}^{\infty} A_{\rm in}(\varpi) \sum_{n=-N}^{N} C_n(\varpi) \me^{-\mi (\varpi + n\Omega) u} \dif \varpi .
\end{equation}
We generate the static reference waveform with the same numerical implementation at $\epsilon=0$, so that it has the same frequency sampling as the moving shell waveform. The shell is then fixed at $R_0$, and we denote the static transmission amplitude in Eq.~\eqref{eq:static_scattering_basis_section} by $C_{\rm stat}(\varpi)$. In this limit,
\begin{equation}\label{eq:floquet_static_limit}
\lim_{\epsilon \to 0} C_0(\varpi) = C_{\rm stat}(\varpi), \quad \lim_{\epsilon \to 0} C_{n\neq0}(\varpi) = 0 .
\end{equation}

\begin{figure}[!htb]
    \centering
    \begin{subfigure}[t]{0.485\textwidth}
         \centering
         \includegraphics[width=\textwidth]{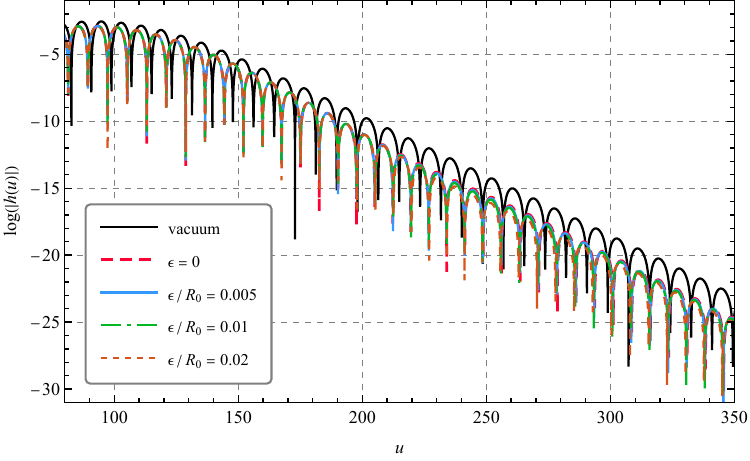}
         \caption{Polar waveform $|h_\epsilon(u)|$.}
         \label{fig:dynamic_polar_waveform}
     \end{subfigure}
     \hfill
     \begin{subfigure}[t]{0.485\textwidth}
         \centering
         \includegraphics[width=\textwidth]{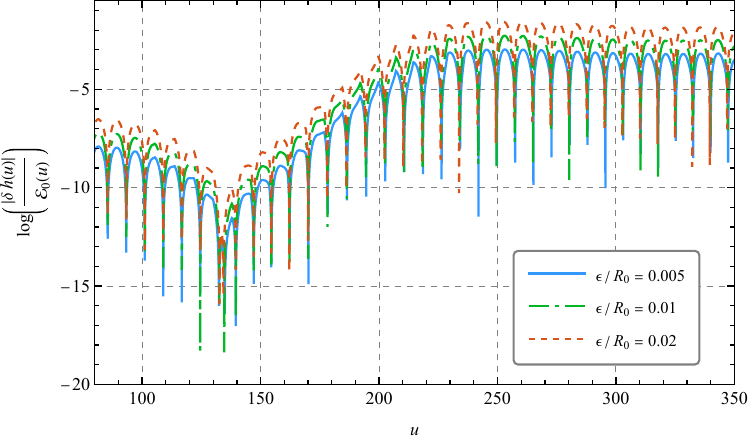}
         \caption{Polar correction $\mathcal{R}_\epsilon(u)$ due to shell motion.}
         \label{fig:dynamic_polar_residual}
     \end{subfigure}
     \hfill
     \begin{subfigure}[t]{0.485\textwidth}
         \centering
         \includegraphics[width=\textwidth]{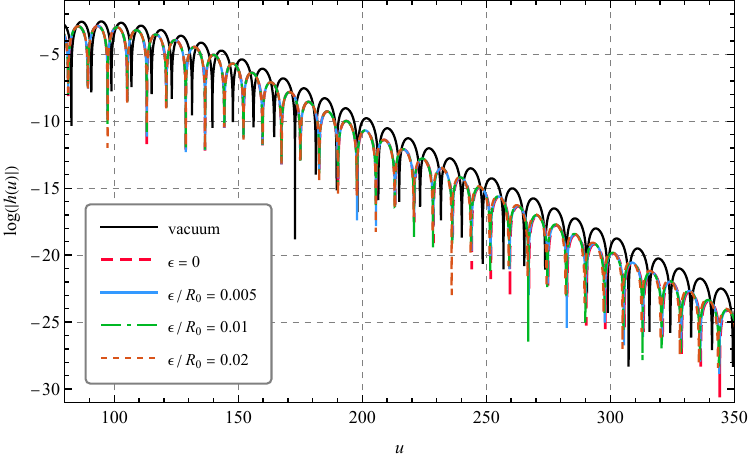}
         \caption{Axial waveform $|h_\epsilon(u)|$.}
         \label{fig:dynamic_odd_waveform}
     \end{subfigure}
     \hfill
     \begin{subfigure}[t]{0.485\textwidth}
         \centering
         \includegraphics[width=\textwidth]{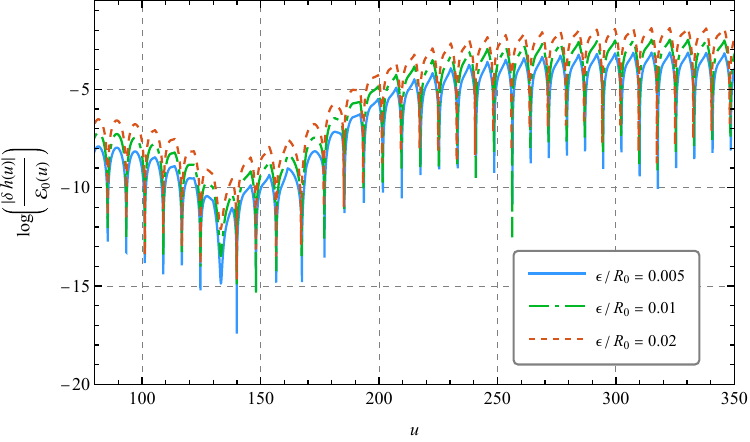}
         \caption{Axial correction $\mathcal{R}_\epsilon(u)$ due to shell motion.}
         \label{fig:dynamic_odd_residual}
     \end{subfigure}
    \caption{Time domain Floquet waveforms of the periodically moving thin shell and the corrections due to its motion.
    Panels (a) and (c) compare vacuum propagation, the static shell reference generated at $\epsilon = 0$ in the Floquet calculation, and moving shell waveforms at $\epsilon/R_0 = 0.005,0.01,0.02$.
    Motion with a small amplitude has little effect on the prompt wave packet.
    Panels (b) and (d) show $\mathcal{R}_\epsilon = |\delta h_\epsilon|/\mathcal{E}_0$, where $\delta h_\epsilon = h_\epsilon-h_0$, $h_0$ is the static reference at $\epsilon=0$, and $\mathcal{E}_0$ is its envelope.
    The correction grows approximately linearly with $\epsilon$ and becomes more visible during late time decay, showing that shell motion mainly modulates the phase and envelope of the late wave train.}
    \label{fig:dynamic_waveforms}
\end{figure}

Fig.~\ref{fig:dynamic_waveforms} shows representative waveforms for $\ell = 2$, $R_0 = 30\,M_+$, $\delta M/M_+ = 0.1$, and $v_s = 0.3$.
Figs.~\ref{fig:dynamic_polar_waveform} and \ref{fig:dynamic_odd_waveform} display $h_\epsilon(u)$ in the polar and axial sectors. Figs.~\ref{fig:dynamic_polar_residual} and \ref{fig:dynamic_odd_residual} show the corrections relative to the static reference at $\epsilon=0$.
To quantify the correction we define
\begin{equation}\label{eq:fixed_R0_motion_residual}
    \delta h_\epsilon(u)=h_\epsilon(u)-h_0(u),\quad
    \mathcal{R}_\epsilon(u)=\frac{|\delta h_\epsilon(u)|}{\mathcal{E}_0(u)},
\end{equation}
where $h_0$ is the static shell waveform at $\epsilon=0$ and $\mathcal{E}_0(u)$ is its envelope.
We use the envelope rather than $|h_0(u)|$ in the denominator to avoid artificial amplification near the nodes of the oscillating static waveform.

Figs.~\ref{fig:dynamic_polar_waveform} and \ref{fig:dynamic_odd_waveform} show the overall waveform scale.
For the small amplitudes $\epsilon/R_0=0.005$, $0.01$, and $0.02$, the moving shell waveforms remain close to the $\epsilon=0$ reference. No additional dominant wave packet appears, and the main peak barely shifts.
Small periodic motion therefore has little effect on the prompt ringdown.
Figs.~\ref{fig:dynamic_polar_residual} and \ref{fig:dynamic_odd_residual} show that $\mathcal{R}_\epsilon$ increases with $\epsilon$ in both sectors, consistent with the first sideband scaling $|C_{\pm 1}|\propto\epsilon$ in the frequency domain.
The correction becomes easier to discern once the static waveform enters its late time decay.

\subsection{Sideband spectra and low frequency shell resonances}

\begin{figure}[!htb]
\centering
\begin{subfigure}[t]{0.485\textwidth}
\centering
\includegraphics[width=\textwidth]{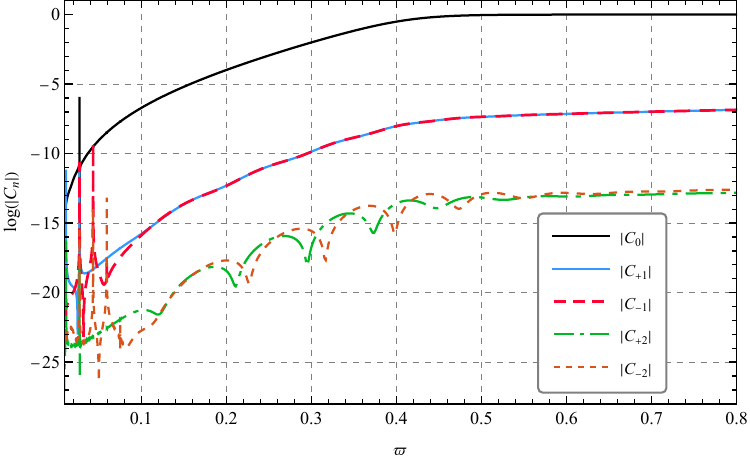}
\caption{Polar sideband spectrum.}
\label{fig:dynamic_polar_sidebands}
\end{subfigure}
\hfill
\begin{subfigure}[t]{0.485\textwidth}
\centering
\includegraphics[width=\textwidth]{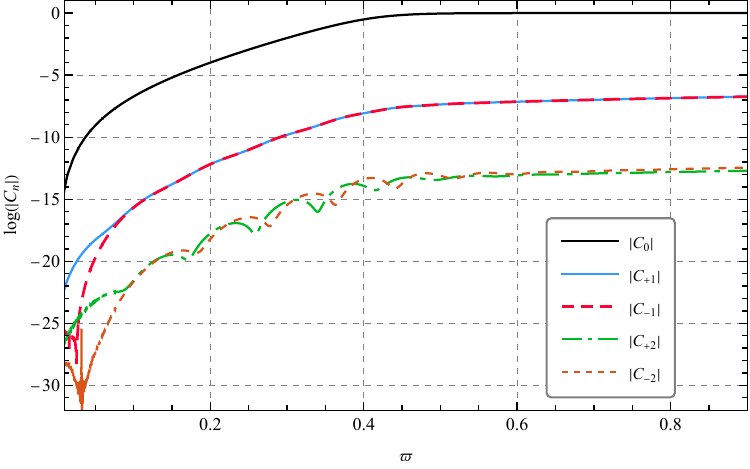}
\caption{Axial sideband spectrum.}
\label{fig:dynamic_odd_sidebands}
\end{subfigure}
\caption{Broadband Floquet sideband spectra $|C_n(\varpi)|$ in the polar and axial sectors at $\epsilon/R_0=0.01$. Only the five output channels $n=0,\pm1,\pm2$ are shown.
The narrow low frequency features are associated with Floquet copies of the static polar shell mode and are examined below.
The broad peak and dip variations of $C_{\pm2}$ at higher frequencies occur in both sectors and reflect coherent contributions from different second order Floquet processes.}
\label{fig:dynamic_sidebands}
\end{figure}

We begin with the sideband spectrum, where frequency conversion at the moving interface can be distinguished from the polar shell mode response.
Fig.~\ref{fig:dynamic_sidebands} shows the broadband Floquet spectrum at $\epsilon/R_0=0.01$.
For a static shell, frequency is conserved and only the central channel $C_0$ survives.
Periodic motion scatters an incident quasifrequency $\varpi$ into the output frequencies $\varpi+n\Omega$, thereby exciting $C_{n\neq0}$.
Figs.~\ref{fig:dynamic_polar_sidebands} and \ref{fig:dynamic_odd_sidebands} show $n=\pm1,\ \pm2$ sidebands in both the polar and axial sectors.
At higher frequencies, $C_{\pm2}$ has broad peaks and dips, while $C_{\pm1}$ varies more smoothly.
The dimensionless separation $\Delta\omega M_+$ of successive features lies between $0.07$ and $0.09$, comparable to the phase scale for a round trip between the photon sphere potential and the shell,
\begin{equation}\label{eq:second_sideband_interference_scale}
    \Delta\omega_{\rm cav} \sim \frac{\pp}{r_{*,+}(R_0)-r_{*,+}(3M_+)} .
\end{equation}
For $R_0=30\,M_+$, Eq.~\eqref{eq:second_sideband_interference_scale} gives $\Delta\omega_{\rm cav}M_+=0.093$, of the same order as the numerical spacing.
The broad features arise from coherent contributions to the second sidebands.
Periodic changes in the shell position and junction conditions generate both direct second order modulation at $\pm2\Omega$ and two successive first order conversions at $\pm\Omega$.
The transmission amplitude can thus be written schematically as
\begin{equation}\label{eq:second_sideband_paths}
    C_{\pm 2}(\varpi)\sim\epsilon^2\left[A_{\pm2}^{\rm dir}(\varpi)+A_{\pm2}^{\rm seq}(\varpi)\right],
\end{equation}
where $A_{\pm2}^{\rm dir}$ denotes direct second order modulation, while $A_{\pm2}^{\rm seq}$ denotes conversion in two steps through an intermediate sideband.
Their relative phase varies with $\varpi$.
When the contributions are nearly in phase, $|C_{\pm2}|$ has a peak. When they nearly cancel, it has a dip.
Although the second sidebands are only of order $\epsilon^2$, their interference pattern is clear on a logarithmic scale.
The first sidebands are dominated by first order modulation and show weaker broadband peak and dip variations.

\begin{figure}[!htb]
\centering
\begin{subfigure}[t]{0.485\textwidth}
\centering
\includegraphics[width=\textwidth]{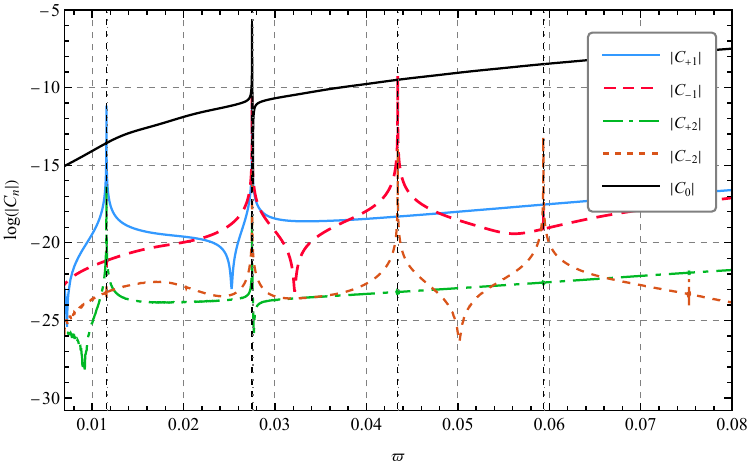}
\caption{Polar low frequency fine scan.}
\label{fig:dynamic_lowfreq_polar_amplitude}
\end{subfigure}
\hfill
\begin{subfigure}[t]{0.485\textwidth}
\centering
\includegraphics[width=\textwidth]{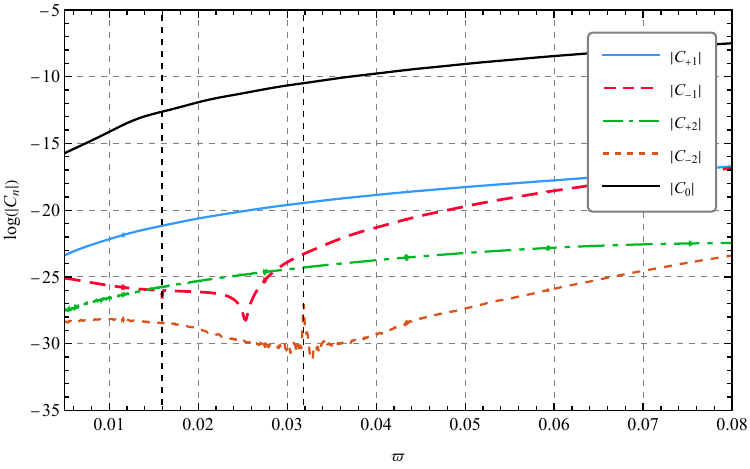}
\caption{Axial low frequency fine scan.}
\label{fig:dynamic_lowfreq_odd_amplitude}
\end{subfigure}
\caption{Fine scan of the low frequency Floquet sidebands.
Panel (a) shows the five polar output channels $|C_n^{\rm polar}(\varpi)|$. Dashed lines mark the reference positions $\varpi_{\rm p}^{(k)}=\omega_{\rm p}-k\Omega$ obtained from the real frequency peak in the central channel.
Narrow peaks appear in several polar channels near these positions, associating them with Floquet copies of the same shell mode.
The pole and zero structure is clearest in $C_0$, whereas zero locations differ among the other channels.
Panel (b) shows the axial sector for comparison.
Its channels have no corresponding peak and dip structure near the polar shell mode references.
The small low frequency features in negative axial sidebands near $\varpi=\Omega$ and $\varpi=2\Omega$ occur where their respective output frequencies $\omega_{-1,+}$ and $\omega_{-2,+}$ pass through zero.}
\label{fig:dynamic_lowfreq_amplitude}
\end{figure}

Several narrow features also appear at the low frequency end of the broadband spectrum.
Fig.~\ref{fig:dynamic_lowfreq_amplitude} resolves these features in the five output channels. They have a different origin from the broad structures at higher frequencies.
In the polar central channel, the low frequency structure consists of a peak and a dip.
Writing their real frequency positions as $\omega_{\rm p}$ and $\omega_{\rm z}$, we find
\begin{equation}\label{eq:pole_zero_position_center}
    \omega_{\rm p} M_+ = 0.027494,\quad \omega_{\rm z} M_+ = 0.027590.
\end{equation}
Shifting these two real frequency features by integer multiples of the shell oscillation frequency gives reference positions for their Floquet copies,
\begin{equation}\label{eq:floquet_pole_zero_copies}
    \varpi_{\rm p}^{(k)}=\omega_{\rm p}-k\Omega,\quad \varpi_{\rm z}^{(k)}=\omega_{\rm z}-k\Omega,
\end{equation}
where $k\in\mathbb{Z}$ labels the sideband order of the copied position.
Both $\varpi_{\rm p}^{(k)}$ and $\varpi_{\rm z}^{(k)}$ refer to real frequency scans.

Fig.~\ref{fig:dynamic_lowfreq_amplitude} shows narrow peaks or enhancements in several polar channels near the same set of $\varpi_{\rm p}^{(k)}$.
The peaks thus reflect a common pole of the Floquet scattering matrix near a polar shell mode, rather than an isolated feature of one output channel.
To describe this locally, we analytically continue the incident frequency $\varpi$ to a complex frequency $\nu$.
Near a pole, the transmission amplitude in an output channel has the approximate form
\begin{equation}\label{eq:local_floquet_pole_form}
    C_n(\nu)\simeq C_n^{\rm reg}(\nu)+\frac{R_n^{(k)}}{\nu-\nu_{\rm p}^{(k)}},
\end{equation}
where $C_n^{\rm reg}$ is the regular part of the transmission amplitude in that channel, and $R_n^{(k)}$ is the residue of the common pole in $C_n$.
The complex pole $\nu_{\rm p}^{(k)}$ follows from the homogeneous Floquet matching conditions and corresponds to the real frequency reference position $\varpi_{\rm p}^{(k)}$.
Its position is set by the Floquet matrix and can therefore appear in several output channels.
If a pole and zero of $C_n(\nu)$ lie close together, denote the nearby zero by $\nu_{{\rm z},n}^{(k)}$.
Its position approximately satisfies
\begin{equation}\label{eq:channel_zero_condition}
    C_n^{\rm reg}\left(\nu_{{\rm z},n}^{(k)}\right)+\frac{R_n^{(k)}}{\nu_{{\rm z},n}^{(k)}-\nu_{\rm p}^{(k)}}=0.
\end{equation}
Unlike the pole, this zero depends on the channel through $C_n^{\rm reg}$ and $R_n^{(k)}$.
The pole and zero structure inherited from the static shell is clearest in $C_0$. The other sidebands usually show sharp spectral peaks and phase jumps instead.
Although $\varpi_{\rm p}^{(2)}=\omega_{\rm p}-2\Omega<0$, the $C_{+2}$ channel still shows narrow peaks near other copied positions.
Near a common matrix pole, sideband coupling excites several components of the output vector according to their respective residues.

The axial sector lacks the corresponding shell matter variables and shows no similar response near the copied frequencies of the polar shell mode.
Indeed, the axial central channel in Fig.~\ref{fig:dynamic_lowfreq_odd_amplitude} remains smooth near $\varpi_{\rm p}^{(k)}$.
Its small low frequency peaks occur mainly when negative sidebands cross zero frequency.
Since $\omega_{n,+}=\varpi+n\Omega$, the $n=-1$ and $n=-2$ channels cross zero frequency at $\varpi=\Omega$ and $\varpi=2\Omega$, respectively, so that
\begin{equation}\label{eq:negative_sideband_thresholds}
    \omega_{-1,+}=0,\quad \omega_{-2,+}=0 .
\end{equation}
For axial perturbations, $h_{0,n}\sim (f/2)(\psi_n+r\partial_r\psi_n)C_n$, whereas $h_{1,n}\sim-\mi\omega_{n,+}r\psi_n C_n/(2f)$.
Thus $h_{0,n}$ remains finite as $\omega_{n,+}\to0$, while $h_{1,n}$ is suppressed by the factor $\omega_{n,+}$.
The axial junction conditions contain $\mathfrak{u}=(\beta/f)h_0+\dot R h_1$ and $\mathfrak{n}=(\dot R/f)h_0+\beta h_1$.
For $n=-1$, the normal condition includes $\beta h_{1,-1}\propto\omega_{-1,+}C_{-1}$ as well as the contribution $[\dot R h_{0,0}/f]_{-1}$ due to shell motion, which remains finite at $\omega_{-1,+}=0$.
As $\varpi$ approaches $\Omega$, $C_{-1}$ must vary rapidly to maintain the junction condition.
Similarly, suppression by $\omega_{-2,+}$ near $\varpi=2\Omega$ causes a rapid change in $C_{-2}$.
These are threshold responses where negative sidebands cross zero frequency. Their positions are set by $\Omega$ rather than by axial eigenmodes.

\begin{figure}[!htb]
\centering
\begin{subfigure}[t]{0.2425\textwidth}
\centering
\includegraphics[width=\textwidth]{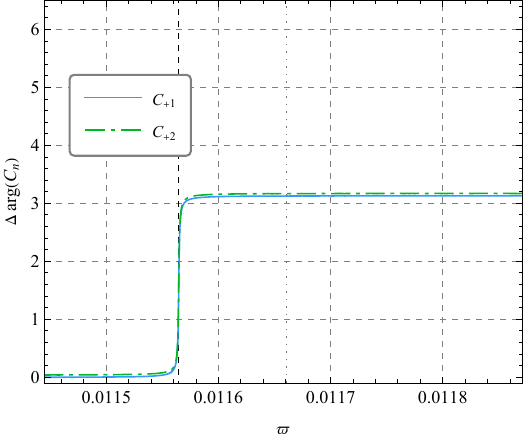}
\caption{$k=1$ window.}
\label{fig:dynamic_phase_kplus1}
\end{subfigure}
\begin{subfigure}[t]{0.2425\textwidth}
\centering
\includegraphics[width=\textwidth]{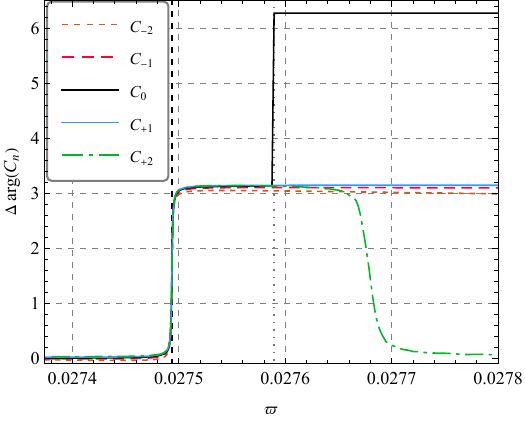}
\caption{$k=0$ window.}
\label{fig:dynamic_phase_k0}
\end{subfigure}
\begin{subfigure}[t]{0.2425\textwidth}
\centering
\includegraphics[width=\textwidth]{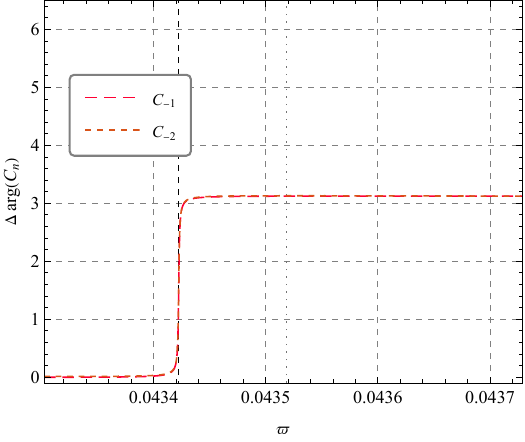}
\caption{$k=-1$ window.}
\label{fig:dynamic_phase_kminus1}
\end{subfigure}
\begin{subfigure}[t]{0.2425\textwidth}
\centering
\includegraphics[width=\textwidth]{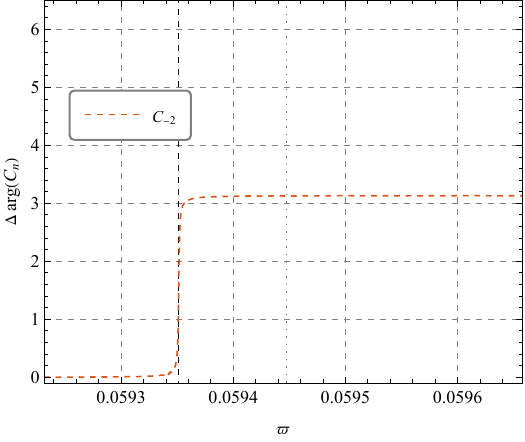}
\caption{$k=-2$ window.}
\label{fig:dynamic_phase_kminus2}
\end{subfigure}
\caption{Local polar phases near narrow low frequency peaks.
Panels (a) through (d) show the $k=1,0,-1,-2$ sideband windows, and the dashed lines mark the real frequency peak copy positions $\varpi_{\rm p}^{(k)}$.
Each panel includes only channels with appreciably enhanced amplitudes and recalculates the phase within the local window.
Noncentral sidebands show jumps of nearly $\pp$ near the peaks, consistent with the same shell mode entering different output channels through sideband coupling.
The nearby zero produces two phase jumps in the central channel $C_0$ in the $k=0$ window.}
\label{fig:dynamic_lowfreq_phase}
\end{figure}

Fig.~\ref{fig:dynamic_lowfreq_phase} examines the local phases near the copied frequencies.
Noncentral sidebands generally show a phase jump close to $\pp$ at the narrow peaks, consistent with a resonance pole close to the real axis.
The nearby zero also affects $C_0$ in the $k=0$ window, giving it a different phase pattern from the other channels.

\begin{figure}[!htb]
\centering
\begin{subfigure}[t]{0.325\textwidth}
\centering
\includegraphics[width=\textwidth]{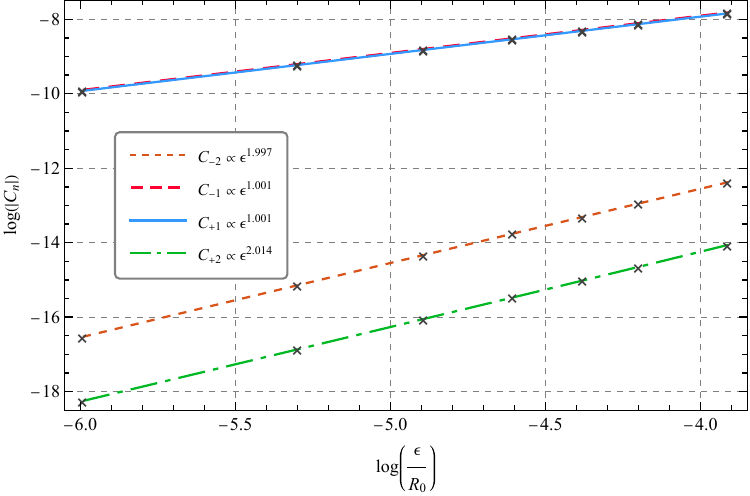}
\caption{Small amplitude scaling.}
\label{fig:dynamic_sideband_scaling}
\end{subfigure}
\hfill
\begin{subfigure}[t]{0.325\textwidth}
\centering
\includegraphics[width=\textwidth]{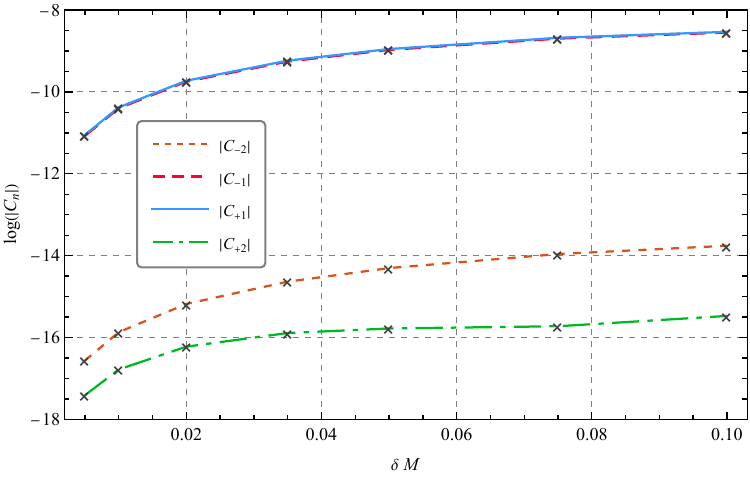}
\caption{Shell mass scan.}
\label{fig:dynamic_mass_scan}
\end{subfigure}
\hfill
\begin{subfigure}[t]{0.325\textwidth}
\centering
\includegraphics[width=\textwidth]{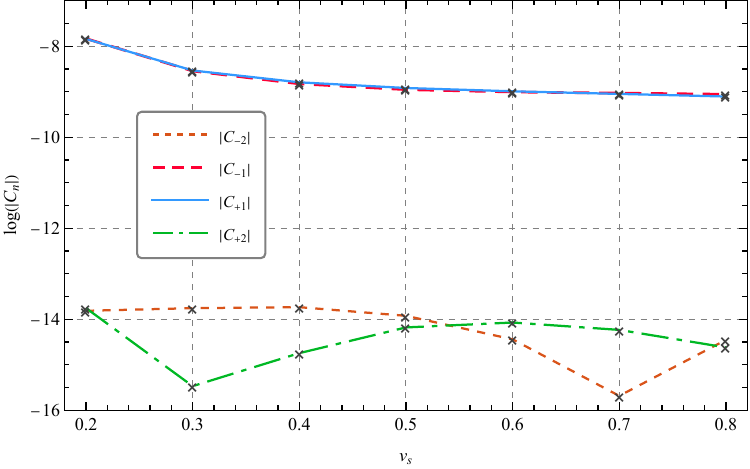}
\caption{Sound speed scan.}
\label{fig:dynamic_sound_speed_scan}
\end{subfigure}
\caption{Amplitude scaling and parameter scans for the dynamical sidebands.
Panel (a) plots each sideband against the oscillation amplitude on logarithmic scales for both axes.
Panels (b) and (c) show the sideband amplitudes at the central frequency as functions of shell mass and surface fluid sound speed.}
\label{fig:dynamic_scaling_scans}
\end{figure}

Fig.~\ref{fig:dynamic_sideband_scaling} shows how the sidebands depend on $\epsilon/R_0$ at fixed $R_0=30\,M_+$.
At small amplitude, the scan gives the scaling expected from perturbation theory
\begin{equation}\label{eq:sideband_epsilon_scaling}
|C_{\pm 1}| \propto \epsilon , \quad |C_{\pm 2}| \propto \epsilon^2.
\end{equation}
These powers agree with perturbation theory at small amplitude and support the interpretation of the sidebands as products of shell motion.
Scans over several parameters also show that macroscopic shell properties affect the sideband strengths.
In Fig.~\ref{fig:dynamic_mass_scan}, decreasing the additional shell mass $\delta M$ reduces the shell's surface stresses and suppresses the noncentral sidebands as the vacuum limit is approached.
Fig.~\ref{fig:dynamic_sound_speed_scan} shows that changing the surface fluid sound speed $v_s$ alters the propagation time of local pressure waves and hence the resonance strength in the polar sidebands.
Axial matching involves neither shell deformation nor pressure fluctuations, so its equations are independent of $v_s$.

\subsection{Excitation of long lived shell modes}

The transmission spectrum contains sidebands and long lived modes, but their appearance in the waveform depends on how strongly they are excited.
We first examine the late time echoes from a distant shell, then consider a closer shell where a long lived polar mode can be excited resonantly.
Echoes are not separate initial excitations. They are secondary signals from a packet that reflects between the shell and the photon sphere potential barrier before leaking through the barrier to infinity. If $r_{\rm peak}$ denotes the peak of the effective potential, their propagation delay is of order $\Delta u_{\rm echo}\simeq2|r_{*,+}(R_0)-r_{*,+}(r_{\rm peak})|$. Reflection phase and barrier filtering can also shift the peaks of the actual wave packet.

In the $R_0=30 M_+$ waveforms of Figs.~\ref{fig:dynamic_polar_waveform} and \ref{fig:dynamic_odd_waveform}, the prompt ringdowns nearly overlap in the polar and axial sectors. The effect of shell motion becomes visible only at late times.
To examine the frequencies present in the late time waveform, we extract complex frequencies from the polar signal with the matrix pencil method (MPM), varying the time window and matrix rank to check the stability of the extraction.
Tab.~\ref{tab:R30_late_mpm_modes} lists the QNM frequencies extracted from the prompt ringdown and late time signal.
The real parts of the static and moving shell prompt ringdown frequencies differ by about $5.0 \times 10^{-4} M_+^{-1}$.
At late times, we identify three matched pairs of modes. Their real frequencies for the moving shell are shifted by $2.8 \times 10^{-3}$ to $6.7 \times 10^{-3} M_+^{-1}$ relative to the static shell.
Their damping rates remain similar, consistent with the late time waveform differences in Fig.~\ref{fig:dynamic_polar_waveform}.

\begin{table}[!htb]
    \centering
    \caption{Complex frequencies extracted from polar waveforms at $R_0=30\,M_+$ by MPM. The last column gives $\Delta\Re(\omega M_+)$, the difference between the real frequencies of the moving and static shells.
    The moving shell has $\epsilon/R_0=0.02$. The other parameters are those of Fig.~\ref{fig:dynamic_waveforms}.
    We use three overlapping windows within $116 \leq u/M_+\leq 152$ for the prompt ringdown and five windows within $150\leq u/M_+\leq 380$ for the late time components, and compare different matrix ranks.}
    \label{tab:R30_late_mpm_modes}
    \begin{tabular}{c c c c}
        \hline
        Waveform component & Static $\omega M_+$ & Moving $\omega M_+$ & $\Delta\Re(\omega M_+)$ \\
        \hline
        Prompt ringdown & $0.368259-0.087480\mi$ & $0.368763-0.088512\mi$ & $0.000504$ \\
        Late time component 1 & $0.288064-0.111884\mi$ & $0.292186-0.110208\mi$ & $0.004122$ \\
        Late time component 2 & $0.373302-0.103775\mi$ & $0.380043-0.103411\mi$ & $0.006741$ \\
        Late time component 3 & $0.422703-0.092144\mi$ & $0.425533-0.095279\mi$ & $0.002831$ \\
        \hline
    \end{tabular}
\end{table}

The late time components in Tab.~\ref{tab:R30_late_mpm_modes} have damping rates of order $10^{-1}M_+^{-1}$, and the waveforms in Fig.~\ref{fig:dynamic_waveforms} show no sustained shell oscillation.
A stronger late time response requires a mode with weaker damping that the incident packet can excite.
We vary the shell radius, mass difference, and sound speed to trace such a mode in the static polar spectrum.

\begin{figure}[!htb]
    \centering
    \includegraphics[width=0.6\textwidth]{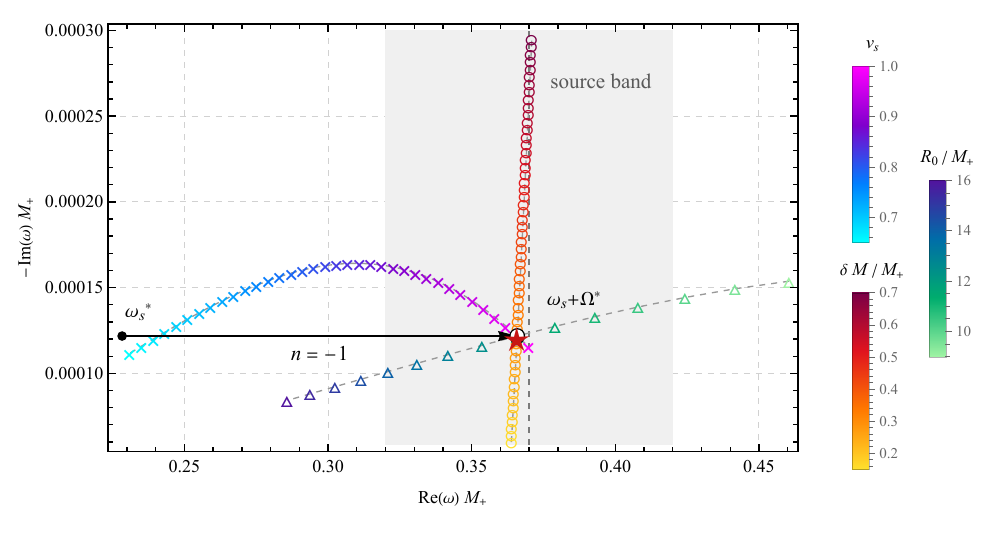}
    \caption{Replica frequency $\zeta_{-1}=\omega_s+\Omega$ of the long lived polar shell mode on the incident side.
    The horizontal and vertical axes are $\Re(\zeta_{-1}M_+)$ and $-\Im(\zeta_{-1}M_+)$, respectively.
    Crosses show the scan in $v_s$ at fixed $R_0=12\,M_+$ and $\delta M/M_+=0.3$. Circles show the scan in $\delta M/M_+$ at fixed $R_0=12\,M_+$ and $v_s=0.99$. Triangles show the scan in $R_0/M_+$ at fixed $\delta M/M_+=0.3$ and $v_s=0.99$.
    The shaded region marks the incident band $\omega_c-\sigma_\omega \leq \Re \omega \leq \omega_c + \sigma_\omega$, and the vertical dashed line marks $\omega_c M_+=0.37$.
    The black dot and star mark the static shell mode $\omega_s$ and its replica $\omega_s+\Omega$ at $R_0=12\,M_+$, $\delta M/M_+=0.3$, and $v_s=0.99$.}
    \label{fig:stage3_shell_replica_scan}
\end{figure}

Periodic motion opens a frequency conversion channel through which the incident wave can excite this mode.
From Eq.~\eqref{eq:floquet_sideband_frequency}, when the $n=-1$ sideband approaches the shell mode frequency $\omega_s$, the corresponding incident frequency is $\varpi \simeq \omega_s+\Omega$.
More generally, we define the shell mode's replica frequency on the incident side as $\zeta_n=\omega_s-n\Omega$ and measure its distance from the center of the incident Gaussian spectrum by
\begin{equation}\label{eq:stage3_source_detuning}
    d_n=\frac{|\Re\zeta_n-\omega_c|}{\sigma_\omega}.
\end{equation}
Fig.~\ref{fig:stage3_shell_replica_scan} shows three parameter scans in the complex frequency plane.
In each scan we vary one of $v_s$, $\delta M/M_+$, and $R_0/M_+$ while holding the other two fixed. We recalculate the shell oscillation frequency $\Omega$ for each parameter set.
As $R_0$ increases, $\Re(\zeta_{-1}M_+)$ moves from above the source band to below it and approaches the spectral center near $R_0=12 M_+$.
Changing the shell mass $\delta M$ has a smaller effect on the real part of the replica frequency, while the damping rate increases with $\delta M$.
We therefore choose $R_0=12 M_+$ and $\delta M/M_+=0.3$ for the waveform calculation. At these values, the shell mode remains weakly damped and its first replica lies within the main part of the incident band.

\begin{figure}[!htb]
    \centering
    \includegraphics[width=0.6\textwidth]{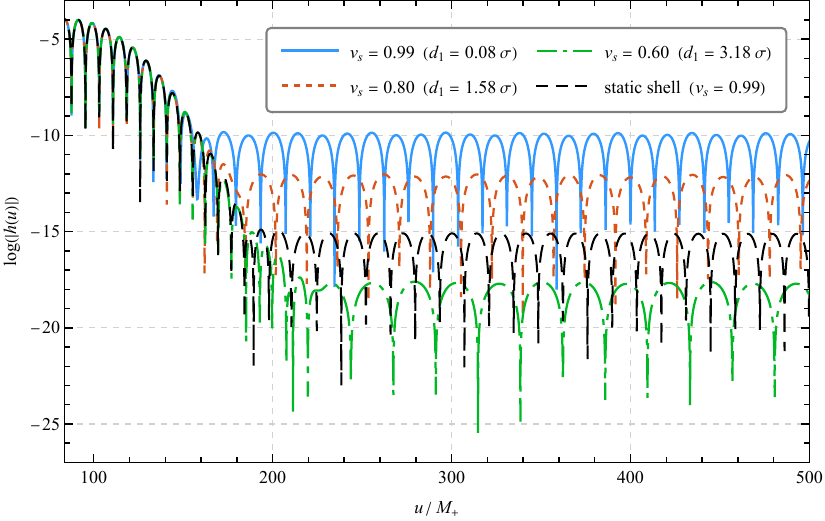}
    \caption{Real polar waveforms at $R_0=12\,M_+$ and $\delta M/M_+=0.3$.
    The three moving shells have $\epsilon/R_0=0.02$, and the static reference has $v_s=0.99$.
    All curves use the same incident packet and normalization. Their main differences appear in the long lived oscillation at late times.}
    \label{fig:stage3_resonant_waveforms}
\end{figure}

For $R_0=12 M_+$, $\delta M/M_+=0.3$, and $v_s=0.99$, the static shell mode has $\omega_s M_+ \simeq 0.228306-1.22 \times 10^{-4}\mi$ and the shell oscillates at $\Omega M_+=0.137545$.
Hence $\Re(\zeta_{-1}M_+) \simeq 0.36585$, only $0.083 \sigma_\omega$ from the center of the incident spectrum.
The static shell mode itself lies on the low frequency side of that spectrum, but motion lets incident radiation near the spectral center excite it through the $n=-1$ channel.
Fig.~\ref{fig:stage3_resonant_waveforms} shows the polar waveforms for $v_s=0.99,\ 0.80,\ 0.60$.
At $v_s=0.80$ and $0.60$, the corresponding values of $d_{-1}$ rise to about $1.58$ and $3.18$.
The first replica thus moves out of the main incident band as $v_s$ decreases.
Higher order replicas may still lie near the spectral center, but their amplitudes are suppressed by higher powers of $\epsilon$.
The prompt ringdown is nearly identical in all these waveforms. Appreciable differences develop only later.
The moving shell produces a sustained late time oscillation at $v_s=0.99$, a weaker signal at $v_s=0.80$, and a still smaller one at $v_s=0.60$.
Over $300<u/M_+<800$, the moving shell waveform at $v_s=0.99$ has an amplitude about $174$ times that of its static counterpart.

To identify the modes that dominate the late time oscillation, we again apply the MPM to the time domain waveforms.
In Tab.~\ref{tab:stage3_floquet_mpm_modes}, the static waveform at $v_s=0.99$ contains a long lived shell mode, whereas the moving shell waveform contains a set of components with nearly equal damping rates and frequencies separated by approximately $\Omega$. The moving shell components satisfy
\begin{equation}\label{eq:stage3_floquet_mode_ladder}
    \omega_q\simeq\nu+q\Omega,\quad -\Im\omega_q\simeq-\Im\nu,
\end{equation}
where $q$ labels the output harmonics of the same Floquet mode.
For $v_s=0.99$, the frequencies of the $q=-1,0,+1$ components are separated by about $0.137545\,M_+^{-1}$ and all have damping rates near $1.20\times10^{-4}M_+^{-1}$.
Equally spaced components are also present at the other two sound speeds, but with much smaller waveform amplitudes.

\begin{table}[!htb]
    \centering
    \caption{Long lived components extracted from late time polar waveforms at $R_0=12\,M_+$.
    The moving shells have $\epsilon/R_0=0.02$. The static row gives the shell mode at $v_s=0.99$.
    The output harmonic is labeled by $q$, and $|A_q|$ is the fitted amplitude referred to $u_*=300\,M_+$.
    The extraction uses $300<u/M_+<800$ and neighboring time windows.}
    \label{tab:stage3_floquet_mpm_modes}
    \begin{tabular}{c c c c c}
        \hline
        Case & $q$ & $\Re(\omega_qM_+)$ & $-\Im(\omega_qM_+)$ & $|A_q|$ \\
        \hline
        Static, $v_s=0.99$ & --- & $0.228306$ & $1.22\times10^{-4}$ & $1.393\times10^{-7}$ \\
        Moving, $v_s=0.99$ & $-1$ & $0.090714$ & $1.20\times10^{-4}$ & $1.348\times10^{-6}$ \\
                              & $0$ & $0.228259$ & $1.20\times10^{-4}$ & $2.429\times10^{-5}$ \\
                              & $+1$ & $0.365804$ & $1.20\times10^{-4}$ & $5.687\times10^{-7}$ \\
        Moving, $v_s=0.80$ & $-1$ & $0.073043$ & $1.60\times10^{-4}$ & $1.101\times10^{-7}$ \\
                              & $0$ & $0.182103$ & $1.59\times10^{-4}$ & $2.790\times10^{-6}$ \\
                              & $+1$ & $0.291162$ & $1.59\times10^{-4}$ & $9.467\times10^{-8}$ \\
        Moving, $v_s=0.60$ & $-1$ & $0.054498$ & $9.1\times10^{-5}$ & $3.016\times10^{-10}$ \\
                              & $0$ & $0.132646$ & $9.0\times10^{-5}$ & $1.073\times10^{-8}$ \\
                              & $+1$ & $0.210796$ & $9.0\times10^{-5}$ & $2.843\times10^{-10}$ \\
        \hline
    \end{tabular}
\end{table}
\FloatBarrier

Tab.~\ref{tab:stage3_floquet_mpm_modes} also shows that at $v_s=0.99$ the central harmonic has an amplitude of $2.429\times10^{-5}$, about $174$ times the amplitude of the corresponding static shell mode, in agreement with the late time ratio of the two waveforms.
Their damping rates are both close to $1.2 \times 10^{-4}M_+^{-1}$.
The enhancement therefore follows from the overlap of the incident spectrum with the first replica, rather than from an appreciable change in the shell mode lifetime.
After subtracting this set of long lived components from the late time waveform, we can still extract shorter lived modes that are also present for a static shell.

\subsection{Exceptional points and resonant waveform fitting}

The late time signal in the preceding subsection arises from resonant excitation of a long lived shell mode.
Shell motion can also couple modes that belong to distinct spectral branches.
When a Floquet replica of one mode approaches the frequency of another, the two complex frequency branches may coalesce.
We return to $R_0=30 M_+$ with $\delta M/M_+=0.1$ to examine this degeneracy and its effect on the waveform.

To determine when two modes coalesce, consider static frequencies satisfying $\omega_a\simeq\omega_b-\Omega$ and solve the homogeneous Floquet problem in the complex frequency plane $\nu$.
Eliminating the other sidebands from $\mathbb{M}^{\mathrm{pol}}(\nu)\mathbb{X}=0$ gives the local root equation for the two nearby modes in the form
\begin{equation}\label{eq:stage4_effective_two_mode}
    \det\!\begin{pmatrix}
        \nu-\widetilde\omega_a & -\kappa_{ab} \\
        -\kappa_{ba} & \nu-(\widetilde\omega_b-\Omega)
    \end{pmatrix}=0,
\end{equation}
where $\widetilde\omega_a$ and $\widetilde\omega_b$ are the two complex frequencies modified by periodic coupling. The explicit $-\Omega$ shifts the replica of the second mode.
The first order periodic couplings $\kappa_{ab}$ and $\kappa_{ba}$ arise from shell motion and are both proportional to $\epsilon$ at small amplitude.
Because perturbations can leak through the horizon and to infinity, these frequencies and couplings are generally complex.
Solving Eq.~\eqref{eq:stage4_effective_two_mode} yields
\begin{equation}\label{eq:stage4_ep_discriminant}
    \nu_\pm=\frac{\widetilde\omega_a+\widetilde\omega_b-\Omega}{2}
    \pm\frac{\sqrt{\mathcal{D}}}{2},\quad
    \mathcal{D}=\left[\widetilde\omega_a-(\widetilde\omega_b-\Omega)\right]^2+4 \kappa_{ab} \kappa_{ba} .
\end{equation}
For $\mathcal{D}=0$, both the real and imaginary parts of this complex discriminant vanish.
If the intermode coupling is nonzero, Eq.~\eqref{eq:stage4_effective_two_mode} then has only one independent eigenvector, giving a second order EP.
The sound speed $v_s$ and relative oscillation amplitude $\epsilon/R_0$ provide two tunable parameters, controlling the detuning of the branches and their sideband coupling.

We locate the critical point using the full polar Floquet matrix.
After eliminating the other components near the two roots, let $F(\nu;v_s,\epsilon)$ denote the resulting local root function.
At coalescence $F_*=\partial_\nu F_*=0$.
Provided $\partial_\nu^2 F_*\neq 0$, an expansion about this point gives
\begin{equation}\label{eq:stage4_ep_local_expansion}
    F \simeq \frac{F_{\nu\nu,*}}{2}(\nu-\nu_*)^2 + F_{v,*}(v_s-v_{s,*})+F_{\epsilon,*}(\epsilon-\epsilon_*),
\end{equation}
where a star denotes evaluation at the EP.
Setting this expression to zero gives the squared separation of the two nearby roots as
\begin{equation}\label{eq:stage4_ep_sqrt_law}
    (\nu_+-\nu_-)^2\simeq -\frac{8}{F_{\nu\nu,*}}\left[F_{v,*}(v_s-v_{s,*})+F_{\epsilon,*}(\epsilon-\epsilon_*)\right] .
\end{equation}
On a slice with $\epsilon=\epsilon_*$, the expected splitting is therefore $|\nu_+-\nu_-|\propto |v_s-v_{s,*}|^{1/2}$.

\begin{figure}[!htb]
    \centering
    \begin{subfigure}[t]{0.525\textwidth}
        \centering
        \includegraphics[width=\textwidth]{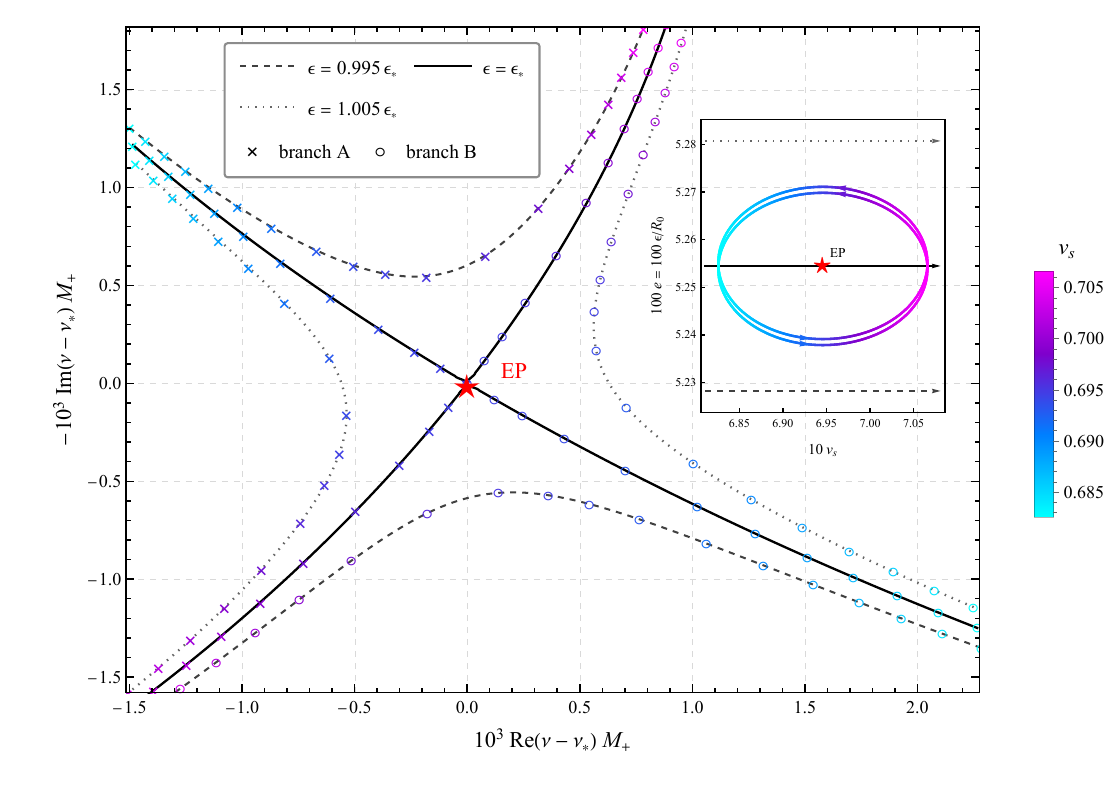}
        \caption{Complex frequency branch trajectories}
        \label{fig:stage4_ep_mode_trajectories}
    \end{subfigure}
    \hfill
    \begin{subfigure}[t]{0.425\textwidth}
        \centering
        \includegraphics[width=\textwidth]{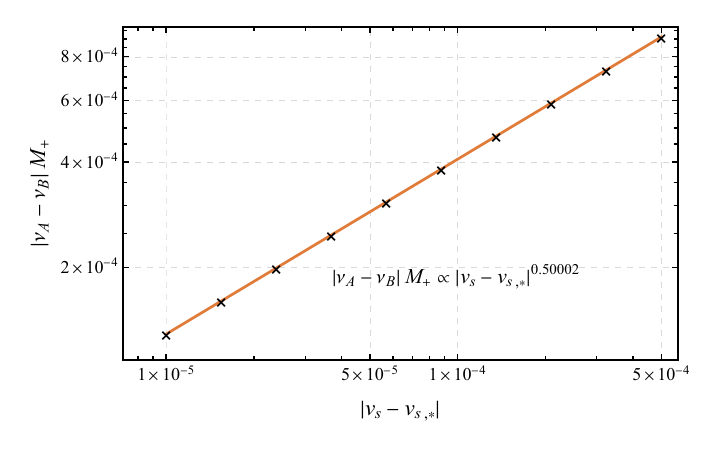}
        \caption{Square root splitting}
        \label{fig:stage4_ep_sqrt_splitting}
    \end{subfigure}
    \par\medskip
    \begin{subfigure}[t]{0.49\textwidth}
        \centering
        \includegraphics[width=\textwidth,trim=20 20 0 45,clip]{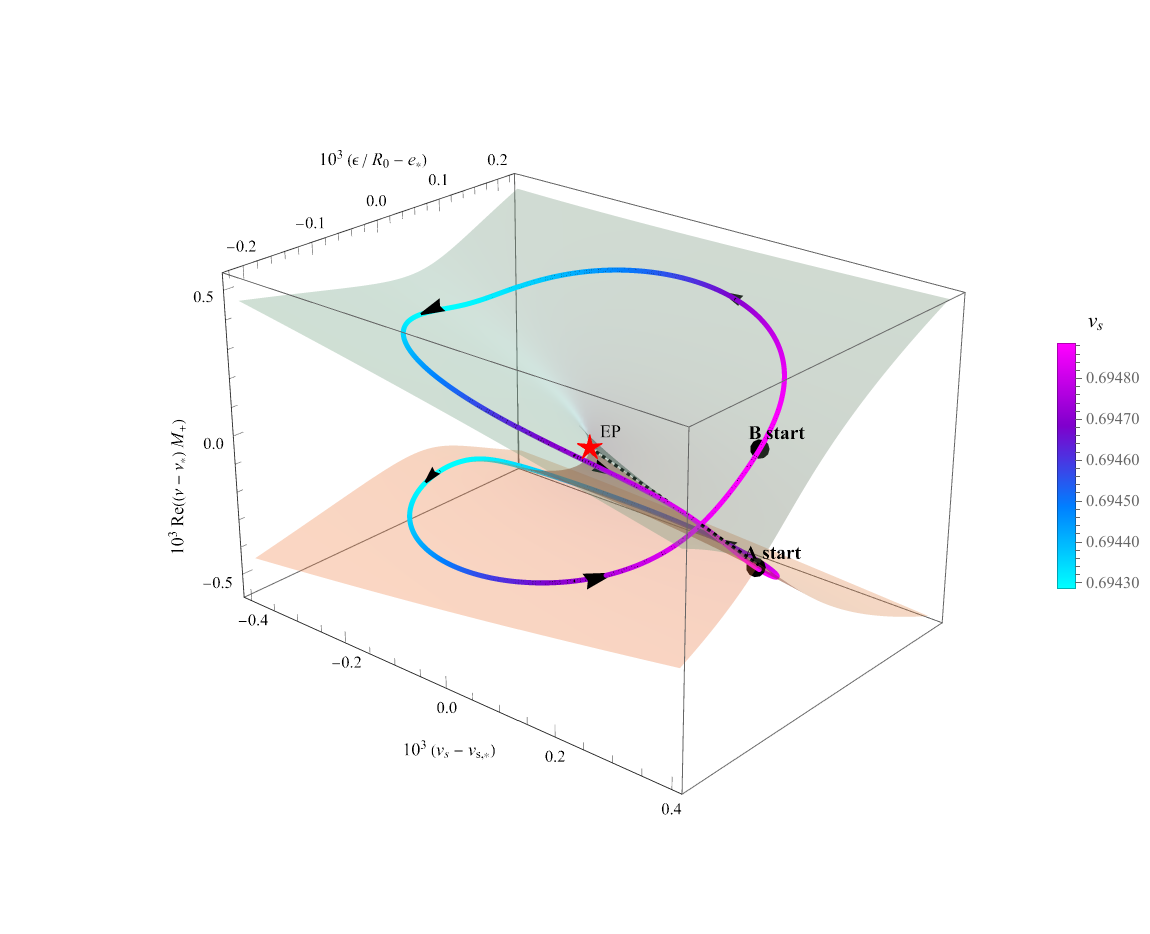}
        \caption{Riemann surface for the real part}
        \label{fig:stage4_ep_riemann_re}
    \end{subfigure}
    \hfill
    \begin{subfigure}[t]{0.49\textwidth}
        \centering
        \includegraphics[width=\textwidth,trim=20 20 0 45,clip]{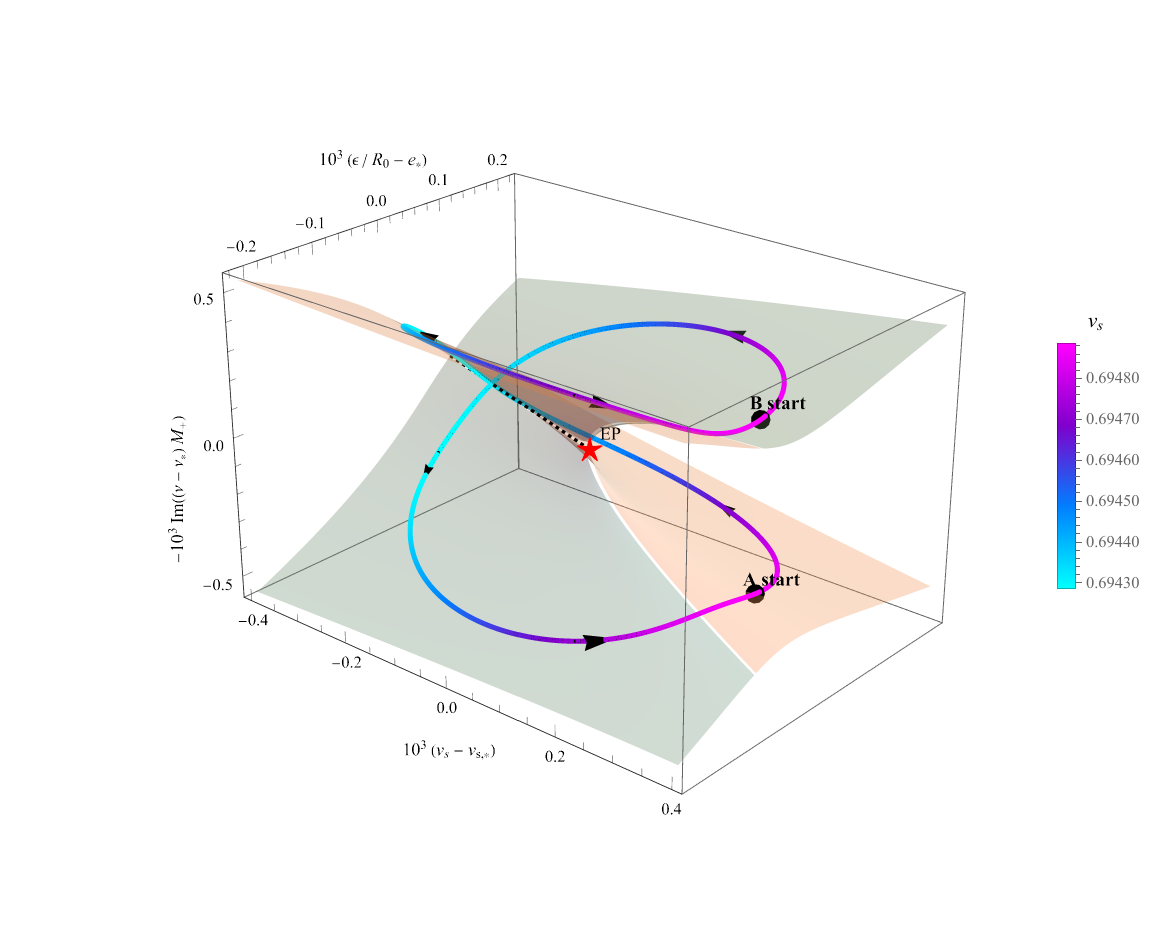}
        \caption{Riemann surface for the negative imaginary part}
        \label{fig:stage4_ep_riemann_im}
    \end{subfigure}
    \caption{Polar Floquet EP at $R_0=30\,M_+$ and $\delta M/M_+=0.1$.
    Panel (a) shows the trajectories of two complex frequency branches as the sound speed varies at $\epsilon/\epsilon_*=0.995,\ 1,\ 1.005$. Its inset shows the scan paths in parameter space and the path that winds twice around the EP. Panel (b) shows the square root splitting of their frequency separation with $|v_s-v_{s,*}|$ at $\epsilon=\epsilon_*$. In panels (c) and (d), the surfaces show the square root approximation, and the colored curves trace the mode trajectories around the EP. Black circular markers indicate the starting points, arrows show the direction of traversal, and black dashed lines denote the branch cuts.}
    \label{fig:stage4_ep_spectrum}
\end{figure}

Solving $F=0$ and $\partial_\nu F=0$ for these parameters gives
\begin{equation}\label{eq:stage4_ep_location}
    v_{s,*}=0.69458625,\quad \epsilon_*/R_0=0.05254489,
    \quad \nu_*M_+=0.37673076-0.09853550\mi .
\end{equation}
At these values, the matching matrix has only one independent eigenvector, and the two complex frequencies and their eigenfunctions coalesce.
Fig.~\ref{fig:stage4_ep_mode_trajectories} traces the two nearby frequency branches as the sound speed varies, for $\epsilon/\epsilon_*=0.995,\ 1,\ 1.005$.
Only at the critical amplitude do the trajectories meet at the EP in Eq.~\eqref{eq:stage4_ep_location}.
The inset of Fig.~\ref{fig:stage4_ep_mode_trajectories} shows three scan paths at fixed amplitude in the $(v_s,\epsilon/R_0)$ plane and a closed path that winds twice around the EP.
At fixed $\epsilon=\epsilon_*$, a logarithmic fit to the frequency separation in Fig.~\ref{fig:stage4_ep_sqrt_splitting} gives an exponent of $0.50002$, consistent with Eq.~\eqref{eq:stage4_ep_sqrt_law}.
Figs.~\ref{fig:stage4_ep_riemann_re} and \ref{fig:stage4_ep_riemann_im} show the local square root Riemann surfaces described by Eq.~\eqref{eq:stage4_ep_local_expansion}. Their vertical axes give the real and negative imaginary parts of the complex frequency offset, respectively. The colored curves trace the corresponding components of the QNM frequencies computed along the closed path. The two roots exchange branches after one circuit and return to their starting branches after two.

\begin{figure}[!htb]
    \centering
    \begin{subfigure}[t]{0.325\textwidth}
        \centering
        \includegraphics[width=\textwidth]{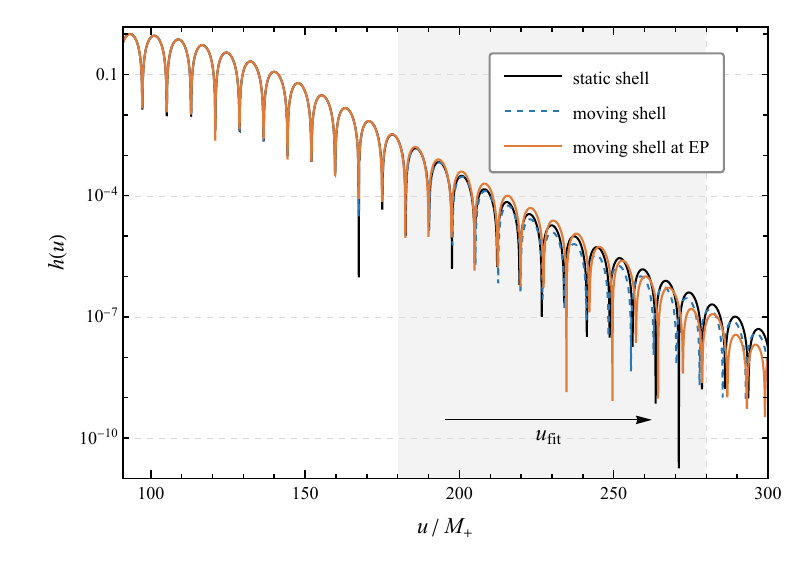}
        \caption{Three polar waveforms}
        \label{fig:stage4_ep_waveforms}
    \end{subfigure}
    \hfill
    \begin{subfigure}[t]{0.325\textwidth}
        \centering
        \includegraphics[width=\textwidth]{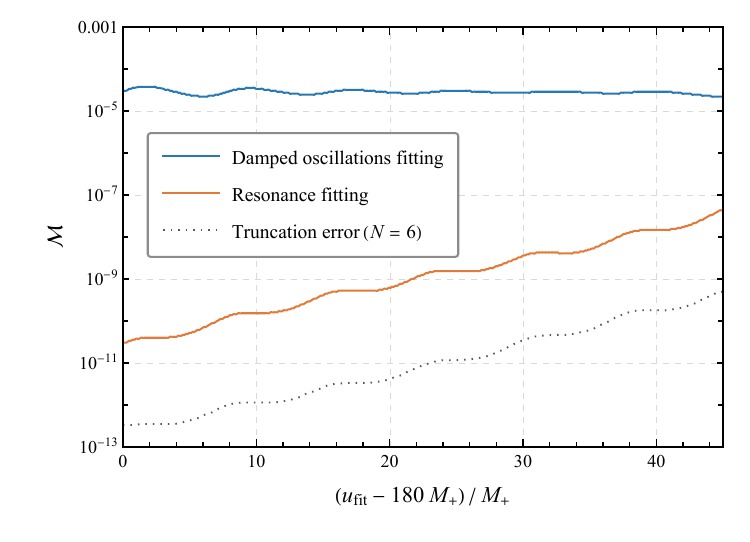}
        \caption{Fit start time and mismatch}
        \label{fig:stage4_ep_fit_start}
    \end{subfigure}
    \hfill
    \begin{subfigure}[t]{0.325\textwidth}
        \centering
        \includegraphics[width=\textwidth]{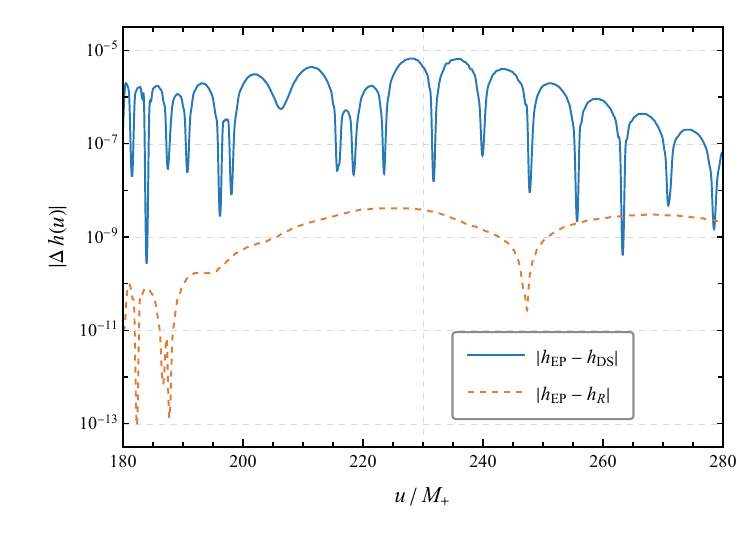}
        \caption{Waveform residuals and resonant fit}
        \label{fig:stage4_ep_fit_residuals}
    \end{subfigure}
    \caption{Polar waveforms and reconstruction at the EP for $R_0=30\,M_+$ and $\delta M/M_+=0.1$.
    Panel (a) compares waveforms for static and ordinarily moving shells with the waveform at the EP, using the same incident packet.
    The static and ordinarily moving shells have $v_s=0.3$, with $\epsilon/R_0=0.02$ for the moving shell. The parameters at the EP are given in Eq.~\eqref{eq:stage4_ep_location}.
    The shaded region marks the fitting window $180\leq u/M_+\leq 280$.
    Panel (b) varies the fit start time at fixed end time $u/M_+=280$. It compares the mismatches of the ordinary damped mode and resonant templates, together with the difference between the $N=6$ and $N=7$ waveforms.
    Panel (c) compares the fitted residuals over $180\leq u/M_+\leq 280$ and fits the resonant component left by the ordinary damped modes at later times.}
    \label{fig:stage4_ep_fits}
\end{figure}

At the EP, the coalescence of the two modes produces a second order pole of the transmission amplitude $C_n(\nu)$ near $\nu_*$.
An incident spectral component at $\nu+m\Omega$ can reach the output channel at $\nu+n\Omega$ through sideband order $n-m$.
Summing these contributions, the transmitted spectral amplitude near $\nu_*$ takes the form
\begin{equation}\label{eq:stage4_ep_laurent}
    \sum_m A_{\rm in}(\nu+m\Omega)C_{n-m}(\nu+m\Omega)
    \sim \frac{b_{2n}}{(\nu-\nu_*)^2}+\frac{b_{1n}}{\nu-\nu_*}+\mathcal{O}(1) .
\end{equation}
The coefficients $b_{2n}$ and $b_{1n}$ are obtained by contour integration around $\nu_*$ in the complex frequency plane.
The inverse Fourier transform of the singular part of Eq.~\eqref{eq:stage4_ep_laurent} gives
\begin{equation}\label{eq:stage4_resonant_profile}
    P_{\rm R}(u)=-2\pp\mi\sum_{n=-N}^{N}\bigl(b_{1n}-\mi u b_{2n}\bigr)
    \me^{-\mi(\nu_*+n\Omega)u} .
\end{equation}
The second order pole thus contributes a term linear in time to every output harmonic, while $\Im\nu_*<0$ ensures that these harmonics still decay.
Fig.~\ref{fig:stage4_ep_waveforms} compares three polar waveforms generated by the same incident packet.
The static and ordinarily moving shells have $v_s=0.3$. For the moving shell, $\epsilon/R_0=0.02$. The shell at the EP has the sound speed and amplitude given in Eq.~\eqref{eq:stage4_ep_location}.
The shell radius and mass difference are the same in all three cases.
Their prompt ringdowns nearly coincide, and differences emerge only at late times.
We therefore use $180 \leq u/M_+ \leq 280$ for the waveform reconstruction and fits.
Define $P_0(u)=-2\pp\mi\sum_n b_{2n}\me^{-\mi(\nu_*+n\Omega)u}$, and let $H_j(u)$ denote the time domain waveforms of four other ordinary damped modes obtained from the frequency domain calculation.
The two fitting templates are
\begin{equation}\label{eq:stage4_fit_models}
    \begin{aligned}
        h_{\rm DS}(u)&=\Re\!\left[\sum_{j=1}^{4}c_jH_j(u)+c_0P_0(u)\right],\\
        h_{\rm R}(u)&=\Re\!\left[\sum_{j=1}^{4}\widetilde c_jH_j(u)
        +\widetilde c_0 P_0(u)+c_{\rm R}P_{\rm R}(u)\right].
    \end{aligned}
\end{equation}
The overall complex amplitudes in these expressions are fitted over the chosen time window.
We quantify the fit by the mismatch
\begin{equation}\label{eq:stage4_fit_mismatch}
    \mathcal{M}(W)=
    \frac{\int_W\bigl[h_{\rm EP}(u)-h_{\rm fit}(u)\bigr]^2\dif u}
    {\int_W h_{\rm EP}^2(u)\dif u},
\end{equation}
where $h_{\rm EP}$ is the numerical waveform at the EP and $h_{\rm fit}$ is its reconstruction from a fitted template.

We first attempt to infer the complex mode frequencies directly from the numerical waveform.
Let $K$ be the number of ordinary damped modes and treat their frequencies, together with the frequency of the resonant term containing the linear time factor, as fitting parameters.
Even with $K=12$, the fitted frequencies vary appreciably with the time window.
The inferred damping rate of the resonant term is about $0.082M_+^{-1}$, compared with $0.0985M_+^{-1}$ from the frequency domain calculation.
Each mode of a periodically moving shell contributes a set of Floquet sidebands that also overlap with contributions from other modes, making their frequencies difficult to identify from the total waveform alone.
We therefore fix $\Omega$ and calculate the sideband weights from the frequency domain residues and incident spectrum. We fit the complex mode frequencies and overall complex amplitudes to the waveform.
For $K=2$, the three complex frequencies and four complex amplitudes give $14$ real fitting parameters. The fitted frequency at the EP differs from Eq.~\eqref{eq:stage4_ep_location} by about $3.0 \times 10^{-4}$.
Increasing $K$ further does not appreciably improve the result.

We next use the four ordinary damped modes in Eq.~\eqref{eq:stage4_fit_models}, fixing their complex frequencies to the frequency domain values and fitting only the overall amplitudes of the waveform templates.
Fig.~\ref{fig:stage4_ep_fit_start} shows the mismatch as the fit start time varies, with the end fixed at $u/M_+=280$.
Over $180\leq u/M_+ \leq 280$, the ordinary damped mode template and the template containing the resonant term yield mismatches of $2.9221\times10^{-5}$ and $2.7602\times10^{-11}$, respectively.
Fig.~\ref{fig:stage4_ep_fit_residuals} compares each reconstructed waveform with the numerical signal.
The ordinary damped modes leave a regularly oscillating residual, which becomes much smaller when the resonant term is included.
Both the windowed mismatch and the pointwise waveform residual thus show the contribution of the second order pole to the late time signal.

\section{Conclusion and discussion}\label{sec:conclusion}

We have studied gravitational perturbations of a BH surrounded by a moving thin shell.
The shell's four velocity and normal change with time, so its junction conditions cannot be obtained by replacing $R_0$ with $R(t)$ in the static conditions. Using a comoving worldtube gauge and a Gaussian decomposition adapted to the shell, we derived the linearized Israel conditions on a general spherically symmetric background \cite{Pani:2009ss,Laeuger:2025zgb}.
Periodic shell motion turns monochromatic scattering into a multichannel Floquet problem. The moving boundary converts frequencies in both the polar and axial sectors, and the shell matter degree of freedom allows polar shell modes to contribute to the sideband response.
In the time domain, a distant moving shell mainly modifies the echoes already present for a static shell, whereas a closer shell can strongly excite a long lived polar shell mode through frequency conversion.
We also verified a second order EP in the polar spectrum and identified the resonant term generated by its second order pole in the late time waveform.

The usefulness of a static shell approximation therefore depends on the part of the waveform being analyzed.
The prompt ringdown is much stronger than the late time echoes, so changes in echo phase and envelope can leave the full waveform nearly unchanged.
For $R_0=12\,M_+$, $\delta M/M_+=0.3$, and $v_s=0.99$, a constant detector noise spectrum gives a signal to noise ratio for the moving shell waveform about $174$ times that of the static shell waveform over $300<u/M_+<800$. The ratio is close to unity when the prompt ringdown is included.
To assess whether the late time enhancement can be detected, one must choose echo search windows using a realistic noise curve and examine whether changes in the central BH parameters are degenerate with shell motion in the waveform.
Existing echo searches show that the choice of window and template affects the statistical interpretation of late time signals \cite{Wang:2019szm,Lo:2018sep}. The bias from fitting a moving shell signal with a static shell template warrants a separate assessment.

The EP found here need not be the only one in this model.
Periodic shell motion produces replicas of every complex mode frequency separated by integer multiples of $\Omega$. Other pairs of branches may therefore coalesce through sideband coupling at different parameters. Similar branch migration near EPs has been studied for Kerr BHs \cite{Cavalcante:2024swt,Motohashi:2024fwt}.
With more than two tunable parameters, EPs may form continuous curves in a larger parameter space. Such curves have been found in perturbed BH QNM spectra \cite{Cao:2025afs}.
Three branches could also coalesce at a higher order EP. For example, the inverse transform of a third order pole contains a resonant term proportional to $u^2$.
Whether any such point appears in a waveform depends on the excitation of the relevant modes and the interference among their contributions, making it necessary to examine both complex frequency poles and the time domain response \cite{Oshita:2025ibu,Yang:2025dbn}.

A next step is to extend the junction conditions on the moving worldtube to rotating backgrounds.
In Kerr spacetime, the radial and angular perturbation equations are coupled. The Floquet effect of shell motion may therefore occur alongside coupling between angular modes. A slow rotation treatment or junction conditions written in Teukolsky variables may provide a route forward \cite{Pani:2012bp,Feng:2024ygo}.
The local barotropic equation of state and nondissipative surface fluid model used here can also be relaxed.
Dissipation, viscosity, or additional matter degrees of freedom would change the responses of the shell displacement $\Phi$ and density perturbation $\delta\Sigma$, and hence the positions of shell modes, Floquet sidebands, and EPs in parameter space.
For shell motion that is not strictly periodic, the range of validity of the discrete sideband approximation remains to be determined.


\section*{Acknowledgement}

This work is supported by the National Natural Science Foundation of China No.~12475067 and No.~12235019.

\bibliographystyle{utphys}
\bibliography{citation}

\end{document}